\documentclass[11pt]{article}
\usepackage[utf8]{inputenc}
\usepackage{authblk}
\usepackage{amsmath}
\usepackage{amsfonts}
\usepackage{graphicx}
\usepackage{setspace}
\usepackage{xcolor}
\usepackage[margin=1in]{geometry}
\usepackage{hyperref}
\usepackage{cleveref}

\title{Morphological evolution of a semiconductor surface driven by irradiation-induced anisotropic plastic flow}

\author[1]{Tyler P. Evans\footnote{Corresponding author: evans.tyler@utah.edu, https://orcid.org/0000-0001-7812-2479}}
\author[2]{Scott A. Norris}
\affil[1]{Department of Mathematics, University of Utah, Salt Lake City, UT 84112, United States of America}
\affil[2]{Department of Mathematics, Southern Methodist University, Dallas, TX 75275, United States of America}

\begin{document}
	
\maketitle

\begin{abstract}
While numerous models exist which explain certain aspects of irradiation-induced nanopatterning on semiconductors, a comprehensive theoretical explanation has remained elusive. However, it is increasingly apparent that such a model will require understanding the dual influence of the collision cascade initiated by ion implantation: first, as a source of material transport by sputtering and atomic displacements occurring over short time scales, and, second, as a source of defects permitting viscous flow within the thin, amorphous layer that results from sustained irradiation over longer time scales. To better understand the latter, we develop several asymptotic approximations for coupling the local ion flux experienced by the amorphous layer to the layer's evolving free interface. From these and the physical hypothesis of irradiation-induced anisotropic plastic flow, or ``ion-hammering", we derive a generalized Kuramoto-Sivashinsky-type equation for the evolving free surface. With physically plausible parameters, the present model achieves good quantitative and qualitative agreement with several aspects of experimental observations of nano-pattern formation during irradiation of silicon by argon, krypton and xenon, and with projectile energies from 500eV to 2keV. Disagreements between model and experiment are discussed, as are implications for future directions.
\end{abstract}

\tableofcontents

\section{Introduction}
Under low energy broad-beam ion irradiation, semiconductor surfaces may spontaneously self-organize into highly regular patterns with characteristic lengths on the order of nanometers \cite{navez-etal-1962,carter-etal-REDS-1977}. The possibility of exploiting this phenomenon to mass-produce nano-engineered surfaces remains appealing \cite{chan-chason-JAP-2007,holmes-cerfon-etal-APL-2012,holmes-cerfon-etal-PRB-2012,munoz-garcia-etal-MSER-2014,NorrisAziz_predictivemodel,cuerno-kim-JAP-2020-perspective}. At the same time, the lack of a fully explanatory theory of this phenomenon continues to pose a decades-old puzzle \cite{chan-chason-JAP-2007,munoz-garcia-etal-MSER-2014,NorrisAziz_predictivemodel,cuerno-kim-JAP-2020-perspective}.

Initial theoretical explanations centered around morphological instability due to surface erosion (i.e., the Bradley-Harper (BH) instability \cite{sigmund-PR-1969,sigmund-JMS-1973,bradley-harper-JVST-1988,bradley-PRB-2011b}), which were soon followed by models considering the redistribution of atoms not sputtered away \cite{carter-vishnyakov-PRB-1996,carter-JAP-1999,norris-etal-2009-JPCM,norris-etal-NCOMM-2011,norris-etal-NIMB-2013,norris-arXiv-2014-pycraters}. Eventually, it was realized that, due to ion-induced structural changes, the first few nanometers of the irradiated semiconductor surface behave as a thin fluid film of extremely high viscosity \cite{rudy-smirnov-NIMB-1999,umbach-etal-PRL-2001}. This provoked a great deal of interest in connecting classical fluid dynamics to the problem of ion-induced self-organization of semiconductor surfaces, with numerous such hydrodynamic-type models treating the amorphous layer and the creation and relaxation of ion-induced stresses as central to pattern formation phenomena \cite{cuerno-etal-NIMB-2011,castro-cuerno-ASS-2012,castro-etal-PRB-2012,norris-PRB-2012-linear-viscous,moreno-barrado-etal-PRB-2015,Swenson_2018,evans-norris-JPCM-2022,evans-norris-JPCM-2023,evans-norris-JEM-2024}, in contrast with models focused on erosion and redistribution. 

Some of these viscous flow models, even if phenomenological, appear to give good agreement with experimental results for at least a subset of possible experimental systems, such as \cite{norris-PRB-2012-linear-viscous,moreno-barrado-etal-PRB-2015}. Increasingly, it seems that a combination of erosion, redistribution, \textit{and} viscous flow effects govern pattern formation, particularly at low energies \cite{norris-etal-SREP-2017,lopezcazalilla-et-al-2018,myint-ludwig-etal-PRB-2021-Ar-bombardment,myint-ludwig-PRB-2021-Kr-bombardment,Teichmann2013}. However, the extent to which hydrodynamic-type effects are responsible for observed nano-patterning remains unclear, and most comparisons of theory and experiment focus on the angle-dependence of nano-ripple formation, whether the ripple wavelength or the critical irradiation angle beyond which ripples begin to form. These predictions are accessible by a linear stability analysis of a proposed model \cite{chandrasekhar-book-2013,cross-greenside-book,NorrisAziz_predictivemodel}, which is suitable to describe morphology evolution at low fluences. Much less attention has been given to directly comparing experimental results with theoretical predictions that require modeling the nonlinear, high-fluence regime of surface morphology evolution, such as the roughness and slope distribution of the irradiated surface whose growth has saturated. In this direction, we develop a continuum model based on the ion-hammering \cite{trinkaus-ryazanov-PRL-1995-viscoelastic,trinkaus-NIMB-1998-viscoelastic,george-etal-JAP-2010}, or anisotropic plastic flow (APF) \cite{norris-PRB-2012-linear-viscous}, effect of broad beam irradiation suitable for comparison with both linear and nonlinear regime observations. Other continuum models of nonlinear regime phenomena have focused on effective body forces \cite{castro-cuerno-ASS-2012,munoz-garcia-etal-PRB-2019} and ``ion winds" \cite{seo-et-al-PRB-2025}.

However, common to all theoretical models is the understanding that the collision cascade initiated by ion implantation is ultimately responsible for the physics underlying pattern formation in irradiated semiconductors, including stress production and relaxation by ion-enhanced fluidity \cite{chan-chason-JAP-2007,liedke-thesis-2011,munoz-garcia-etal-MSER-2014,NorrisAziz_predictivemodel,cuerno-kim-JAP-2020-perspective}, and even phase change \cite{hofsass-bobes-zhang-JAP-2016,evans-norris-JEM-2024}. Naturally, the total ion flux directed at the irradiated surface is widely understood as an important experimental parameter in all theoretical models. As irradiation drives surface evolution, the local ion flux experienced by a particular region within the amorphous layer varies with the surface geometry, in part due to \textit{geometric flux dilution} \cite{norris-etal-SREP-2017}. This nonuniform ion flux is expected to lead to a spatial distribution of stress and fluidity \cite{kalyanasundaram-AM-2006,chan-chason-JVSTA-2008,moreno-barrado-etal-PRB-2015,munoz-garcia-etal-PRB-2019,evans-heyen-PRE-2026}, producing a multilateral coupling between the influence of the collision cascade, the free interface, and the amorphous layer's fluid-like dynamics. Accounting for this nonuniformity constitutes a large portion of our model development.

This paper is structured as follows. First, we introduce a continuum model of the irradiated semiconductor as a viscous fluid undergoing APF, which has been studied elsewhere as a plausible explanation for pattern formation. Then, hypothesizing that the local strength of APF should vary according to the local ion flux, we develop asymptotic approximations for the local ion flux in several cases of interest. These approximations are valid for \textit{any} hydrodynamic model, under mild assumptions, and also lead to a generalization of the description of the amorphous-crystalline interface given by \cite{evans-norris-JEM-2024}. We then derive a generalized Kuramoto-Sivashinsky (gKS) equation for the evolution of the irradiated free surface. Although gKS-type equations are expected generically for irradiated semiconductors \cite{seo-et-al-PRB-2025}, the advantage of the present work is that the gKS's coefficients are fully determined by the ion distribution, which can be obtained from simulations, and two model parameters which can be estimated by experiments \cite{madi-thesis-2011,ishii-etal-JMR-2014,perkinsonthesis2017,NorrisAziz_predictivemodel}. Since gKS equations generically form patterns \cite{NorrisAziz_predictivemodel}, a gKS can be tuned to reproduce experimental observations while bypassing mechanism identification.

When our model is equipped with BCA-informed ion implantation parameters and two other model parameters close to values obtained elsewhere, analysis and simulation of the resulting gKS for several ion-target-energy combinations reveal highly favorable quantitative comparison with experimental observations in both the linear and nonlinear regimes of morphology evolution. Where quantitative accuracy cannot be explored, qualitative model properties are instead compared, such as angle and flux dependence. Last, implications for the alignment of theory and experiment, including plausible explanations for discrepancies, and several future directions are discussed.

\section{Model}
As ions implant into an irradiated surface, they produce flow defects \cite{kinchin-pease-1955,volkert-JAP-1991,ishii-etal-JMR-2014} which eventually lead a previously crystalline substrate to become amorphous and capable of viscous flow \cite{volkert-JAP-1991,umbach-etal-PRL-2001,chan-chason-JAP-2007,chan-chason-JVSTA-2008}. The extent of damage sufficient to cause amorphization is typically on the order of a few nanometers for low-energy irradiation \cite{chan-chason-JAP-2007,munoz-garcia-etal-MSER-2014,wesch-wendler-book-2016,NorrisAziz_predictivemodel,cuerno-kim-JAP-2020-perspective}, leading to the formation of an amorphous-crystalline interface \cite{belyakov-titov-REDS-1996,titov-etal-NIMB-2003,titov-etal-AVPS-2003,titov-etal-PRB-2006}, which acts as the lower boundary of a thin fluid film.

We adopt this well-established characterization of the amorphous thin film as a fluid of very high viscosity. Following existing convention and empirical observations \cite{van-dillen-etal-APL-2001-colloidal-ellipsoids,van-dillen-etal-APL-2003-colloidal-ellipsoids,van-dillen-etal-PRB-2005-viscoelastic-model,vandillen-etal-prb-2006}, we hypothesize that the main effect of the ion beam is to induce anisotropic plastic flow (APF) \cite{otani-etal-JAP-2006,norris-PRB-2012-linear-viscous}, also known as ion-hammering \cite{trinkaus-ryazanov-PRL-1995-viscoelastic,trinkaus-NIMB-1998-viscoelastic}. The full continuum model departs from that of \cite{norris-PRB-2012-linear-viscous} in that (i) we will explicitly consider the local flux dependence of APF and (ii) the shape of the amorphous-crystalline interface is approximated directly from BCA-informed ion implantation parameters \cite{ziegler-biersack-littmark-1985-SRIM}. Our respective analyses then proceed along substantially different lines.

\paragraph{Bulk equations.} Within the amorphous layer, conservation of mass is applied as
\begin{equation}
    \nabla\cdot \vec{v} = 0, \label{eqncombined}
\end{equation}
where $\vec{v}$ is the Eulerian velocity field. Conservation of momentum is written as
\begin{equation}
    \nabla \cdot \textbf{T} = 0,
\end{equation}
the small-Reynolds number limit of $\rho \left( \frac{\partial \vec{v}}{\partial t} + \vec{v} \cdot \nabla \vec{v} \right) = \nabla \cdot \textbf{T}$, where $\textbf{T}$ is the Cauchy stress tensor and $\rho$ is density. Following discussion in \cite{norris-PRB-2012-linear-viscous}, we use the Cauchy stress tensor
 \begin{equation}
 			\textbf{T} = - p\textbf{I} + 2\eta(\dot{\textbf{E}} - \tau_{D}\dot{\textbf{E}}_b),
        \end{equation}
with $\dot{\textbf{E}} = \frac{1}{2}\left( \nabla \vec{v} + \nabla \vec{v}^T \right)$, the linear strain rate tensor. In the above, $p$ is pressure, $\eta$ is the viscosity (here, assumed constant), and $\textbf{I}$ is the identity tensor. APF \cite{van-dillen-etal-APL-2001-colloidal-ellipsoids,van-dillen-etal-APL-2003-colloidal-ellipsoids,van-dillen-etal-PRB-2005-viscoelastic-model,vandillen-etal-prb-2006,otani-etal-JAP-2006,norris-PRB-2012-linear-viscous} is incorporated through the extra term
        \begin{equation} \label{tensordef}
			\dot{\textbf{E}}_b = f A_D
			\begin{bmatrix} 
				\frac{3}{2}\cos(2\theta) - \frac{1}{2} & 0 & \frac{3}{2}\sin(2\theta) \\
				0 & 1 & 0 \\
				\frac{3}{2}\sin(2\theta) & 0 & -\frac{3}{2}\cos(2\theta) - \frac{1}{2} \\
			\end{bmatrix}.
	\end{equation}
Above, $A_D$ is the per-ion deformation due to APF, and $\tau_D=\tau_D(x,z)$ is a dimensionless scalar function which imparts spatial inhomogeneity. In Section \ref{section:analysis}, it will be assigned a specific functional form.

\paragraph{Boundary conditions.} 
At the free upper interface, $z=h(x,t)$, kinematic and stress-balance conditions are applied,
	\begin{equation}
		\begin{gathered}
			v_{I} = \vec{v}\cdot \hat{\mathbf{n}}\\
			\textbf{[[T]]} \cdot \hat{\mathbf{n}} = \gamma \kappa \hat{\mathbf{n}},
		\end{gathered}
	\end{equation}
where $v_I$ is the interfacial velocity, $\hat{\mathbf{n}}$ is the outward-normal vector from the surface, $\kappa$ is the mean curvature ($\kappa=\frac{h_{xx}}{(1+h_x^2)^{3/2}}$ in the one-dimensional case), $\gamma$ is the surface energy (see, e.g., \cite{jaccodine-TES-1963}). At the lower, amorphous-crystalline interface, $z=g(x,t)$, we apply the no-penetration condition,
     \begin{equation} 
        \vec{v}\cdot\hat{\mathbf{n}} = 0, \label{no-pen}
    \end{equation}
and the no-slip condition,
    \begin{equation} 
        \vec{v}\cdot\hat{\mathbf{t}} = 0. \label{no-slip}
    \end{equation}
Here, $\hat{\mathbf{t}}$ is the unit tangent vector from the amorphous-crystalline interface. These are simplifications of the boundary conditions considered in \cite{evans-norris-JEM-2024}, where the effect of phase change at the translating amorphous-crystalline boundary was studied. We neglect these effects in the present work to focus on the amorphous-crystalline interface shape and the effect of depth dependence.

\paragraph{Note on model assumptions.} Here, we have \textit{assumed}, on a phenomenological basis, that APF is an adequate description of the influence of the ion beam. While this functional form has been used with apparent success in connecting theory and experimental results \cite{george-etal-JAP-2010,madi-thesis-2011,norris-PRB-2012-linear-viscous,norris-etal-SREP-2017}, its physical motivation in the case of low-energy (nuclear stopping regime \cite{wesch-wendler-book-2016}) bombardment is unclear \cite{norris-PRB-2012-linear-viscous}. We have also explicitly chosen to neglect erosion \cite{bradley-harper-JVST-1988,bradley-PRB-2011b}, redistribution \cite{carter-vishnyakov-PRB-1996}, ion-induced swelling \cite{Swenson_2018,evans-norris-JPCM-2022,evans-norris-JPCM-2023}, and phase change at the amorphous-crystalline interface \cite{evans-norris-JEM-2024}. This leads to a simplified model which nonetheless produces reasonable agreement with certain experimental systems. Moreover, neglect of erosion and redistribution is motivated by an interest (i) in understanding surface evolution in the case of negligible-sputtering and (ii) in exploring how much of irradiation-induced nanopatterning, even at energies above sputter threshold, can be explained without erosion and redistribution as the dominant mechanisms, aligned with questions raised in, e.g., \cite{cuerno-etal-NIMB-2011,castro-cuerno-ASS-2012,norris-etal-SREP-2017,NorrisAziz_predictivemodel}.

\section{Analysis}
\label{section:analysis}
\begin{figure}[h!]
\begin{center}
\includegraphics[width=.85\linewidth,clip=true,trim=0 0 0 0]{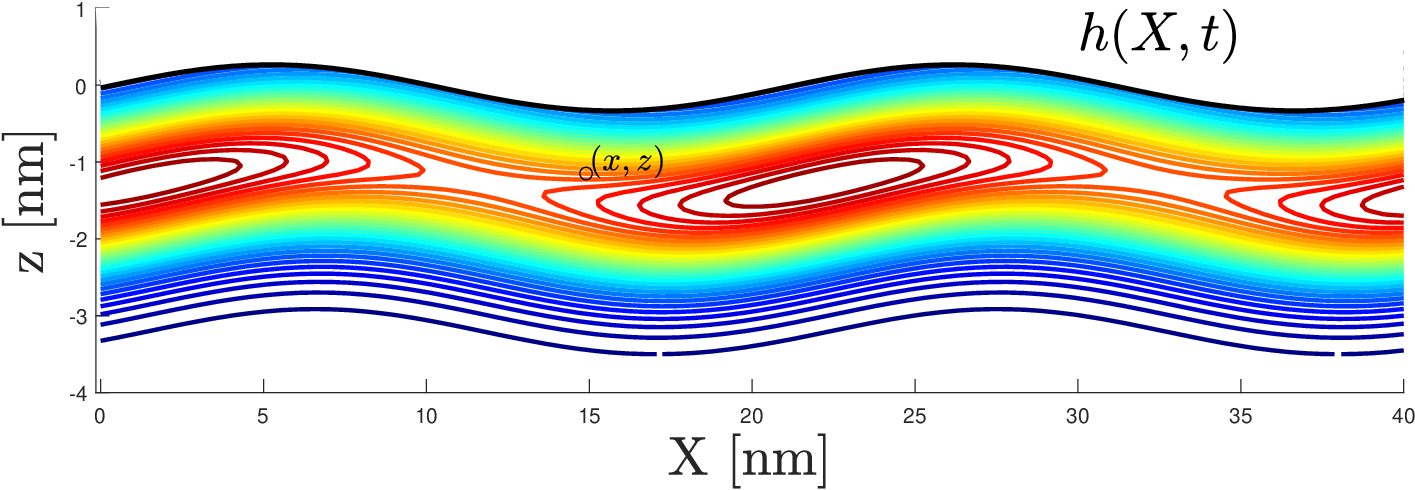}
\end{center}
\caption{Example of level sets of localized flux of ions through a deformed free surface (black curve) at $\theta=45^{\circ}$ off-normal incidence according to Equation (\ref{deposition-integral}). Red regions receive the highest local dose, and purple regions receive the lowest. We have supposed that the free interface is described by $h(X,t) = \frac{3}{10}\sin\big(\frac{3X}{10}\big)$. We choose parameters associated with 250eV Ar$^+$ irradiation of Si: $a=1.8$ nm, $\alpha=0.7$ nm, and $\beta=0.8$ nm \cite{srim-2000.40}. Where the intensity of ion implantation falls off, the amorphous-crystalline interface begins. In Section \ref{section:analysis}, we determine closed-form approximations for (i) the amorphous-crystalline interface and (ii) the share of the ion flux $f$ received by the point $(x,z)$.}
\label{fig:levelcurves}
\end{figure}

The continuum model represented by Equations (\ref{eqncombined})-(\ref{no-slip}) is incomplete without specifying how the amorphous-crystalline interface, $z=g$, is related to the free interface, $z=h$, and without specifying $\tau_D$, which scales the local strength of APF. In this Section, we perform an asymptotic analysis in preparation for equipping the continuum model with a physically-realistic amorphous-crystalline interface and local flux distribution within the amorphous interior.

To this end, we first recall that essentially all existing models of ion-induced pattern formation agree on a key role for the \textit{collision cascade}--- the sequence of atomic recoils induced by a newly implanted ion \cite{kinchin-pease-1955,windischmann-JAP-1987,davis-TSF-1993-simple-compressive-stress,liedke-thesis-2011,wesch-wendler-book-2016}. The main difference between models concerned primarily with erosion and redistribution \cite{bradley-harper-JVST-1988,norris-etal-2009-JPCM,norris-etal-NCOMM-2011,norris-arXiv-2014-pycraters,bradley-PRB-2011b} and those concerned primarily with stress production and relaxation \cite{rudy-smirnov-NIMB-1999,umbach-etal-PRL-2001,chan-chason-JVSTA-2008,norris-PRB-2012-linear-viscous,ishii-etal-JMR-2014} is how the collision cascade is mechanistically linked to pattern formation.

Over many implantations through the same patch of surface, the statistically averaged distribution of a deposited quantity (such as ions or energy) is reasonably well-approximated as a Gaussian ellipsoid \cite{chan-chason-JAP-2007,hossain-etal-JAP-2012,harrison-bradley-PRB-2014,NorrisAziz_predictivemodel}, while this approximation breaks down close to grazing incidence. Deposition can be lessened by \textit{geometric flux dilution} \cite{NorrisAziz_predictivemodel}. Deposition of ions originating from patches of surface that are normal to incoming ions occurs at a greater rate than that originating from patches locally parallel to incoming ions. Hence the influence of ion-implantation on surface evolution is modified by the geometry of the evolving surface. 

For this work, it is sufficient to model the cross-section of the irradiated surface, with $x$ as the projected downbeam coordinate and $z$ depthwise, as is common; see, e.g., \cite{moreno-barrado-etal-PRB-2015,munoz-garcia-etal-PRB-2019}. Restricted to this plane, the integral
\begin{equation}
	D_f(x,z,t) = \int_{-\infty}^{\infty}\frac{\cos(\theta) + h_X\sin(\theta)}{\sqrt{1+h_X^2}} \Psi(x,z;X,h(X,t))dX \label{deposition-integral}
\end{equation}
provides the distribution of ions received at each point $(x,z)$ in the amorphous bulk given uniform irradiation across the surface $h(X,t)$. Ion implantation, recoils, and energy deposition all (at least approximately) follow the form of a Gaussian ellipsoid \cite{sigmund-PR-1969,sigmund-JMS-1973,bradley-harper-JVST-1988,chan-chason-JAP-2007,bradley-PRB-2011b,hossain-etal-JAP-2012,harrison-bradley-PRB-2014,NorrisAziz_predictivemodel},
\begin{equation}
    \Psi(x,z; X, h(X,t)) = \frac{1}{2\pi\alpha \beta}e^{\big(-\frac{[(x-X)\sin(\theta) - (z-h(X,t))\cos(\theta)-a]^2}{2\alpha^2} - \frac{[(x-X)\cos(\theta) + (z-h(X,t))\sin(\theta)]^2}{2\beta^2} \big)}, \label{bigpsi}
\end{equation}
where $a,\alpha,\beta$ are the mean downbeam implantation depth, downbeam standard deviation, and crossbeam standard deviation, respectively \cite{chan-chason-JAP-2007,wesch-wendler-book-2016}. Equation (\ref{bigpsi}) describes the probability of an ion aimed at the surface at location $(X,h(X,t))$ becoming deposited at position $(x,z)$, and its integration in Equation \ref{deposition-integral} leads to the scalar field $D_f(x,z,t)$, visualized for a fixed surface $H(X,\cdot)$ in Figure \ref{fig:levelcurves}. The same form has been used elsewhere in order to study surface erosion rates due to deposited energy, leading to the classical Bradley-Harper (BH) instability \cite{bradley-harper-JVST-1988}. While thorough analyses of Equation \ref{deposition-integral} exist for the erosive-distributive family of models \cite{bradley-harper-JVST-1988,cuerno-barabasi-PRL-1995,bradley-PRB-2011b}, a similar analysis is lacking for the hydrodynamic family of models. Here, as in previous analyses, we make the assumption that ``memory effects" are absent: that modeling the instantaneous flux experienced by the point $(x,z)$ is sufficient, and that we do not need to track the total fluence experienced by $(x,z)$. If this were necessary, one would simply integrate up to the current time $t$. However, doing so would greatly complicate the present analysis, since we would then be required to deal with \textit{nonlinear partial integro-differential equations}.

In what follows, we summarize the results of an asymptotic analysis that are applicable to studying nonlinear surface evolution --- asymptotic approximations of (i) the \textit{local ion flux}, and (ii) a closed-form description of the amorphous-crystalline interface shape --- and then to use them in developing the surface evolution equation. For readability, most calculation details are deferred to the Appendix. 

\paragraph{Small-curvature limit.} We consider Equation (\ref{deposition-integral}) in the limit of small curvature. If the slope of $h(X,t)$ changes slowly enough over any interval that the surface admits a linear approximation, $h(X,t) = h(x,t) + (X-x)h_x$, over that interval, then a small-curvature approximation of $D_f(x,z,t)$ is 
\begin{equation}
    D_f(x,z,t) \approx \frac{1}{\sqrt{2\pi}} \frac{c + h_xs}{\sqrt{1+h_x^2}} \frac{\exp\bigg[-\frac{\big((z-h(x,t) ) + a(c+sh_x)\big)^2 }{2\beta^2(s-ch_x)^2 + 2\alpha^2(c+sh_x)^2} \bigg]}{\sqrt{\beta^2(s-ch_x)^2 + \alpha^2(c+sh_x)^2}}, \label{Dsmallcurvature}
\end{equation}
where $c=\cos(\theta)$ and $s=\sin(\theta)$. Above, $h_x$ denotes $\frac{\partial h}{\partial x}$, the partial derivative of $h(x,t)$ with respect to $x$.

\paragraph{Small-slopes limit.} Here, we note that all explicit $x$-dependence in Equation (\ref{Dsmallcurvature}) has been absorbed into $h(x,t)$. Moreover, if all slopes $h_x$ of interest are small in the laboratory coordinates, then (\ref{Dsmallcurvature}) admits a series expansion in $h_x \approx 0$. We also note that linear approximations of depth-dependence for ion-induced deformation feature in some of the literature \cite{moreno-barrado-etal-PRB-2015,munoz-garcia-etal-PRB-2019,evans-norris-JPCM-2023}. Following this, we record here the linear approximation of Equation (\ref{Dsmallcurvature}) about $z= h(x,t)$, $$D_f(x,z,t) \approx D_{f0}(x,z,t) + D_{f1}(x,z,t)\big(z-h(x,t)\big)$$ where
\begin{equation}
\begin{gathered}
    D_{f0}(x,z,t) = \frac{1}{\sqrt{2\pi}}\frac{c+h_xs}{\sqrt{1+h_x^2}}\frac{\exp\bigg[-\frac{\big(a(c+sh_x)\big)^2 }{2\beta^2(s-ch_x)^2 + 2\alpha^2(c+sh_x)^2} \bigg]}{\sqrt{\beta^2(s-ch_x)^2 + \alpha^2(c+sh_x)^2}} \\
    D_{f1}(x,z,t) = \frac{-a(c+h_xs)D_{f0}(x,z,t)}{2sch_x(\alpha^2-\beta^2) + s^2(h_x^2\alpha^2 + \beta^2) + c^2(h_x^2\beta^2 + \alpha^2 )}. \label{D0-lineardepth}
\end{gathered}
\end{equation}

\begin{figure}[h!]
\begin{center}
\includegraphics[width=1\linewidth]{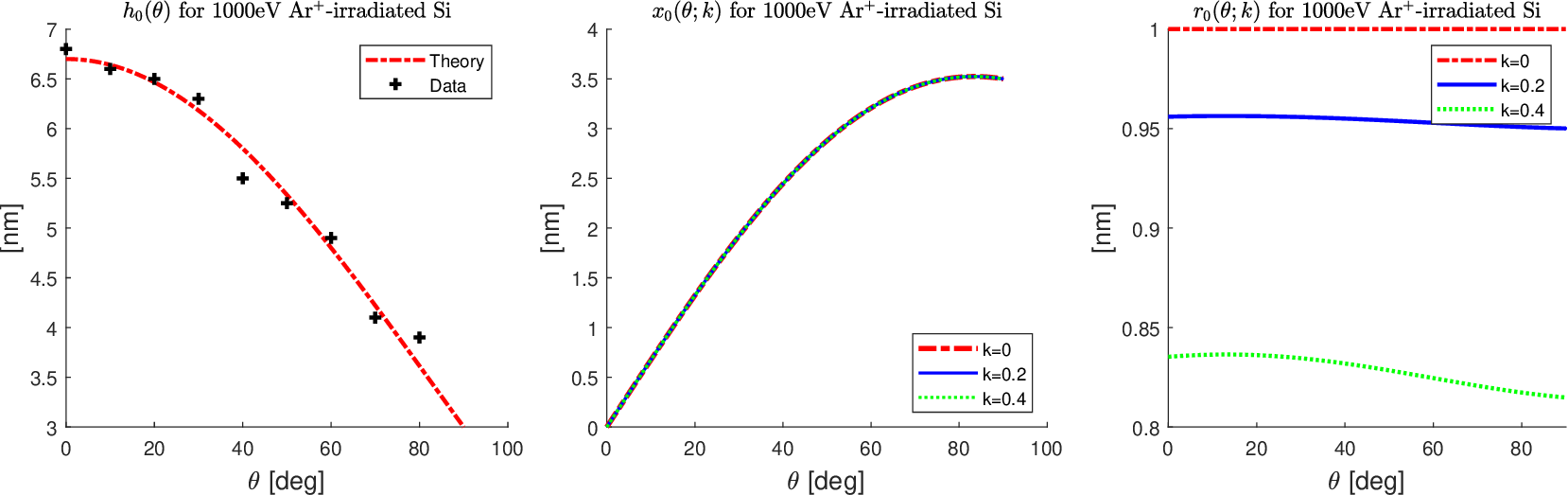}
\end{center}
\caption{Examples of computed $h_0(\theta),x_0(\theta;k)$ and $r_0(\theta;k)$ based on Equations (\ref{interfaces-arbitrary}) for 1000eV Ar$^+$-irradiated Si. \textbf{Left}: comparison of theoretical angle-dependent film thickness with film thickness inferred from experiments \cite{norris-etal-SREP-2017}. \textbf{Center:} the lateral shift separating free and amorphous-crystalline interfaces plotted for three wavenumbers. \textbf{Right:} the flattening factor plotted for three wavenumbers. Note the convergence to 1 for long waves. }
\label{fig:CCbased}
\end{figure}

\paragraph{Location of amorphous-crystalline interface.}
TEM images of irradiated semiconductors frequently reveal an amorphous-crystalline interface that resembles a vertically-translated and horizontally-shifted copy of the free, upper interface \cite{chini-etal-PRB-2003-TEM,ziberi-etal-PRB-2005,moreno-barrado-etal-PRB-2015}. Accordingly, we consider the free interface $z=h(x,t)$ and the amorphous-crystalline interface $z=g(x,t)$ related by
\begin{equation}
    g(x,t) = r_0(\theta;k)h(x-x_0(\theta;k),t) - h_0(\theta).
\end{equation}
In the Appendix, we show that this expression also naturally emerges from an asymptotic analysis of the level curves characterizing how much of the flux $f$ is received by the few nanometers beneath the irradiated free interface. It is also strongly suggested by Figure \ref{fig:levelcurves}, where Equation (\ref{deposition-integral}) is numerically computed as an example. From our analysis, we obtain
\begin{equation}
\begin{gathered}
    h_0(\theta) = ac + 2\sqrt{\alpha^2c^2 + \beta^2s^2}, \\
    x_0(\theta;k) = \frac{1}{k}\arctan\bigg[\frac{\sin(ak\sin(\theta)) + \frac{2k(\alpha^2-\beta^2)sc\cos(ak\sin(\theta))}{\sqrt{\alpha^2c^2+\beta^2s^2} } } {\cos(ak\sin(\theta)) - \frac{2k(\alpha^2-\beta^2)sc\sin(ak\sin(\theta))}{\sqrt{\alpha^2c^2+\beta^2s^2 } } } \bigg],   \\
    r_0(\theta;k) = e^{-\frac{k^2}{4}\big((\alpha^2+\beta^2) - (\alpha^2-\beta^2)\cos(2\theta) \big)} \bigg[1 + \frac{k^2(\alpha^2-\beta^2)^2\sin^2(2\theta)}{\alpha^2c^2 + \beta^2s^2}\bigg], \label{interfaces-arbitrary}
\end{gathered}
\end{equation}
which is shown to be valid to leading order for both small-amplitude perturbations to $h(x,t)$ \textit{and} long-wave perturbations to $h(x,t)$ of arbitrary amplitude. Figure \ref{fig:CCbased} shows $h_0(\theta),x_0(\theta;k)$ and $r_0(\theta;k)$ for 1000eV Ar$^+$-irradiated Si \cite{norris-etal-SREP-2017}. We also note that the derivation presented in the Appendix generalizes \cite{evans-norris-JEM-2024} to small-amplitude perturbations of arbitrary wavenumber $k$, rather than $k \ll 1$ only.

\paragraph{Nonlinear evolution equation.} The stated results above are mathematically general, describing the instantaneous share of the nominal flux $f$ received by a point $(x,z)$ through a deformed interface of small curvature or small slope; they could be adopted into any other existing hydrodynamic-type model \cite{moreno-barrado-etal-PRB-2015,munoz-garcia-etal-PRB-2019}. It remains to connect these results to the nonlinear dynamics of the continuum model represented by Equations (\ref{eqncombined})-(\ref{no-slip}). 

Consistent with the analysis up to this point, observed features on irradiated semiconductor surfaces are typically \textit{long-wave} and of small slope \cite{chini-etal-PRB-2003-TEM,kumar-etal-NRL-2013-preamorphization,moreno-barrado-etal-PRB-2015}. Accordingly, we apply the standard lubrication scalings and nondimensionalization \cite{oron-davis-bankoff-RMP-1997,craster-matar-APS-2009,ajaev-book-2012} to Equations (\ref{eqncombined})-(\ref{no-slip}) as $x = \frac{h_0 \tilde{x}}{\epsilon}, y=\frac{h_0 \tilde{y}}{\epsilon}, z = h_0\tilde{z}, t = \frac{h_0 \tilde{t}}{\epsilon u_0}, u = u_0\tilde{u}, v=u_0\tilde{v}, w=\epsilon u_0 \tilde{w}$, which enable an asymptotic expansion wherein the velocity and pressure fields are adiabatically coupled to the height field. Here, $h_0$ is again the angle-dependent film thickness, $u_0$ is mean lateral flow velocity, and $\epsilon$ is a typical, nondimensionalized wave number, taken to be small.

For conciseness and as a tractable first study, $\tau_D$ is assigned as the further expansion of (\ref{D0-lineardepth}) in small slopes $h_x$. Accordingly, we write
\begin{equation}
    \tau_D(\tilde{z},\tilde{h},\tilde{h}_{\tilde{x}}) = \tau_{00} + \tau_{10}\tilde{h}_{\tilde{x}} + \tau_{01}\cdot(\tilde{z}-\tilde{h}) + \tau_{11}\cdot(\tilde{z}-\tilde{h})\tilde{h}_{\tilde{x}},
\end{equation}
where $\tau_{00},\tau_{10},\tau_{01}$ and $\tau_{11}$ are coefficients involving only $s=\sin(\theta),c=\cos(\theta),a,\alpha,\beta$. In particular, 
\begin{equation}
\begin{gathered}
\tau_{00} = H_1\big(\alpha^2c^2 + \beta^2s^2\big); \hspace{.25cm}
\tau_{01} = H_1a\cos(\theta) \\
\tau_{10} = H_2\bigg[1 - \frac{a^2\beta^2c^2}{(\alpha^2c^2+\beta^2s^2)^2}\bigg]; \hspace{.25cm}
\tau_{11} = -H_2\bigg[\frac{a\beta^2c\big((a^2-2\alpha^2)c^2 - 2\beta^2s^2\big)}{(\alpha^2c^2 + \beta^2s^2)^3}\bigg]
\end{gathered}
\end{equation}
where
\begin{equation}
\begin{gathered}
H_1 = \frac{2\cos(\theta)\exp\bigg[\frac{-a^2}{2(\alpha^2 + \beta^2\tan^2(\theta))}\bigg]}{\sqrt{2\pi}\big(\alpha^2 + \beta^2 + (\alpha^2-\beta^2)\cos(2\theta)\big)}; \hspace{.25cm} 
H_2 = \frac{\sin(\theta)}{\sqrt{2\pi}} \exp\bigg[\frac{-a^2}{2(\alpha^2 + \beta^2\tan^2(\theta))}\bigg].
\end{gathered}
\end{equation}
The application of these lubrication scalings and expansion of (\ref{D0-lineardepth}) results in the nonlinear partial differential equation, 
\begin{equation}
\begin{gathered}
    h_t = C_1h_{xx} + C_2h_xh_{xx} + C_3h_x^2h_{xx} + C_4h_{xx}^2 + C_{5}h_{xxx} + C_{6}h_{x}h_{xxx} \\ + C_{7}h_{xx}h_{xxx} + C_{8}h_{xxxx},
    \label{NLPDE}
\end{gathered}
\end{equation}
a \textit{generalized Kuramoto-Sivashinsky equation} (gKS), where the temporal evolution of free surface $h(x,t)$ is expressed entirely in terms of its first through fourth spatial derivatives, $h_x,h_{xx},h_{xxx},h_{xxxx}$. For readability, we have deferred the exact expression for each coefficient to the Appendix. Here, it suffices to observe that they are fully determined by the parameters $a,\alpha,\beta,\theta,\gamma,f,A_D,\eta$.

While pattern formation occurs generically for gKS equations like (\ref{NLPDE}), the coefficients $C_i$ serve to connect the gKS equation (\ref{NLPDE}) to the physical hypothesis that APF scales locally and instantaneously with the local ion flux, as well as to impart ion-target-energy specificity through the parameters $(a,\alpha,\beta)$. Since $a,\alpha,\beta$ can be obtained via BCA simulations \cite{ziegler-biersack-littmark-1985-SRIM}, $\gamma$ is known to reasonable precision \cite{jaccodine-TES-1963,vauth-mayr-PRB-2007,vauth-mayr-PRB-2008b}, and $f$ and $\theta$ are experimentally prescribed, the only two free parameters in the model are $A_D$ and $\eta$. However, $A_D$ and $\eta$ have also been experimentally estimated for low-energy irradiation of Si \cite{george-etal-JAP-2010,norris-etal-NCOMM-2011,madi-thesis-2011,norris-PRB-2012-linear-viscous,ishii-etal-JMR-2014,evans-norris-JPCM-2023}, providing some basis --- even if tentative --- for the numerical study of Equation (\ref{NLPDE}) featured in Section \ref{section:results}. That $A_D$ and $\eta$ are, in principle, experimentally accessible means that Equation (\ref{NLPDE}) provides specific, testable hypotheses.



Here, we note that this assignment of $\tau_D$ imparts an important physical hypothesis. While we do not know \textit{a priori} that the local strength of APF should vary spatially like the local ion flux, it is perhaps the simplest assignment that is consistent with known ion beam physics. It also appears to be broadly consistent with the spatial dependence of stress found in simulations \cite{kalyanasundaram-AM-2006,moreno-barrado-etal-PRB-2015}.

\paragraph{Linearization in long-wave limit.} Linearizing Equation (\ref{NLPDE}) about $h(x,t) = h_0 + \epsilon e^{\Sigma t + ikx}$ for $\epsilon \ll 1$, we find $\Sigma(k) = -C_1k^2 - C_{5}ik^3 + k^4C_{8}$. Maximizing $\text{Re}(\Sigma)$, we compute the fastest-growing wavenumber $k^* = \sqrt{\frac{C_1}{2 C_{8} }}$. In general, there is no guarantee that $\text{Re}(\Sigma)>0$ for any particular irradiation angle and set of parameters. Rather, one typically observes surfaces that remain flat for irradiation experiments conducted below some critical irradiation angle, $\theta_c$. For irradiation angles $\theta$ above the critical angle $\theta_c$, $\text{Re}(\Sigma)>0$ for at least some $k>0$; then we obtain a theoretical estimate of the amplification factor that should be observed,
\begin{equation}
    \text{Re}\big(\Sigma(k^*)\big) = -\frac{C_1^2}{C_{8}}. \label{amplification_factor}
\end{equation}
One therefore expects to observe $\lambda(\theta) = \frac{2\pi}{k^*}$; that is,
\begin{equation}
\lambda(\theta) = 2\pi\sqrt{\frac{2\gamma h_0(\theta)}{-3fA_D\eta\big(\cos(2\theta)\big(3\tau_{00} +2h_0\tau_{01}\big) - \sin(2\theta)\big(\frac{3}{2}\tau_{10} + h_0\tau_{11} + 3\tau_{00}\frac{x_0}{h_0} + 3\tau_{01}x_0\big) \big)}}. \label{wavelengths}
\end{equation}
This shows that while the ratio $\frac{\gamma}{fA_D\eta}$ scales $\lambda(\theta)$ when $\theta>\theta_c$, $\theta_c$ itself is determined entirely by the parameters $a,\alpha,\beta$ through the terms $\tau_{00},\tau_{01},\tau_{10},\tau_{11}, x_0$ and $h_0$. The \textit{phase velocity} of a single ripple of wavenumber $k$ is expected to be 
\begin{equation}
V_{\text{phase}} = \frac{-\text{Im}(\Sigma(k^*))}{k^*} = C_{5}k^{*2}. \label{velocity}
\end{equation}
Because $C_{5} = fA_D\big[\frac{2}{3}h_0^3\tau_{10} + \frac{3}{4}h_0^4\tau_{11}\cos(2\theta)\big]$
and $h_0,\tau_{10},\tau_{11}$ are defined completely in terms of $(a,\alpha,\beta)$, observing velocities allows an alternative method of measuring $A_D$ for a given system if, e.g., XPCS data is available \cite{myint-ludwig-etal-PRB-2021-Ar-bombardment,myint-ludwig-PRB-2021-Kr-bombardment}. It also provides an even simpler point of comparison with experimental observations: the direction of ripple propagation. This is explored in Section \ref{section:results}.

Finally, we note that two of the ion-induced effects ignored in the present model --- ion-induced swelling (IIS) \cite{Swenson_2018,evans-norris-JPCM-2022,evans-norris-JPCM-2023} and phase change at the amorphous-crystalline interface \cite{evans-norris-JEM-2024} --- do not contribute to $\text{Im}(\Sigma)$. This provides an assurance that an estimate of $A_D$ and $\eta$, and good alignment of the present model with observed ripple velocity, will not be undone by the reintroduction of these neglected mechanisms which properly belong to a more complete model. Moreover, we will show in Section \ref{section:results} that Equation (\ref{wavelengths}) becomes Equation (\ref{wavelengths_withIIS}) when IIS is reintroduced. The effect of reintroducing IIS is simply to translate the wavelengths predicted by Equation (\ref{wavelengths}) to higher values of $\theta$ while leaving the $\theta$-dependence itself essentially unchanged up to translation. A similar modification to Equation (\ref{wavelengths}) would be expected if phase change at the amorphous-crystalline interface were reintroduced \cite{evans-norris-JEM-2024}.

\section{Results}
\label{section:results}
When supplemented with the unknown, but experimentally measurable, parameters $fA_D$ and $\eta$, the present model makes predictions of growth rates (amplification factor), saturated roughness, time-evolving wavelength, ripple velocity, and, informed by \cite{ishii-etal-JMR-2014,evans-heyen-PRE-2026}, flux dependence. In this Section, we present a comparison of experimental observations of these aspects of surface evolution with model predictions. These results are organized as follows. First, we consider wavelength predictions made by Equation (\ref{wavelengths}) against experimental data 500eV Ar$^+$-irradiated Si and 500eV Xe$^+$-irradiated Si \cite{moreno-barrado-etal-PRB-2015}. Second, we consider wavelength predictions from Equation (\ref{wavelengths}) and observations of flux-dependent amplification factor $\Sigma$ of \cite{ishii-thesis-2013} for 600eV Ar$^+$-irradiated Si. Third, we compare theoretical predictions of ripple velocity, (\ref{velocity}), against the XPCS experiments of \cite{myint-ludwig-PRB-2021-Kr-bombardment} and wavelengths from \cite{myint-ludwig-PRB-2021-Kr-bombardment,seo-et-al-JPCM-2022} for 1keV Kr$^+$-irradiated Si. Finally, we compare the results of numerically integrating Equation (\ref{NLPDE}) with parameters corresponding to 2000eV Kr$^+$-irradiated Si with the roughening and wavelength time series of \cite{engler-etal-PRB-2014}. 

\paragraph{Notes on theoretical-experimental comparison and parameter choices.} For lack of experimental estimates of $fA_D$ and $\eta$ for these systems, and motivated by discussion in \cite{mayr-et-al-PRL-2003} where it is suggested that ion-induced viscous flow varies only weakly with energy and perhaps across ion species, we will use $\gamma = 1.36$ J/m$^2$ \cite{vauth-mayr-PRB-2007,vauth-mayr-PRB-2008b}, and $fA_D = 3 \times 10^{-4}$ 1/s and $\eta = 150$ GPa s, the values estimated in \cite{norris-etal-SREP-2017} for 1000eV Ar$^+$-irradiated Si at a flux of 0.02 ions/nm$^2$/s, unless otherwise stated. Lacking parameter estimates for all the systems that we would like to study, the next best thing is, perhaps, to avoid deviating too much from these estimates. This avoids the descent of the present work into a speculative parameter fitting exercise.

Naturally, the present model does not perfectly predict experimentally-observed $\theta_c$. While it is well-established experimentally that $\lambda(\theta)$ diverges as $|\theta - \theta_c|^{-1/2}$ (see, e.g., \cite{moreno-barrado-etal-PRB-2015}) for low fluence, it is not obvious that the surface morphology, or morphology evolution, at high fluence similarly depends on $|\theta-\theta_c|$. Here, we adopt the physical hypothesis that the nonlinear, high fluence regime of surface evolution, like the linear, low fluence regime, depends on $|\theta-\theta_c|$. With this hypothesis in mind, we will sometimes compare dynamics at a prescribed $\theta$ for both the linear and nonlinear regimes based on the distance of $\theta$ from $\theta_c$ predicted by the model even if this is different than experimental $\theta$. We will find that this style of comparison appears to be reasonably well-justified.

Last, we note that, while other experimental observations exist for low-energy irradiated Ge targets \cite{Teichmann2013,perkinson-JVSTA-2013-Kr-Ge}, we make no effort to compare with those observations due to heightened uncertainty surrounding $fA_D, \eta$, and the possible presence of ion-induced swelling (IIS) \cite{Swenson_2018,evans-norris-JPCM-2022,evans-norris-JPCM-2023} in such systems. Including IIS here would require the use of coupled nonlinear partial differential equations, which we wish to avoid in a relatively simple first study.

\subsection{500eV Ar\texorpdfstring{$^+$}{+}-irradiated Si and Xe\texorpdfstring{$^+$}{+}-irradiated Si: angle-dependent wavelengths}
\begin{figure}[h!]
\begin{center}
\includegraphics[width=1\linewidth]{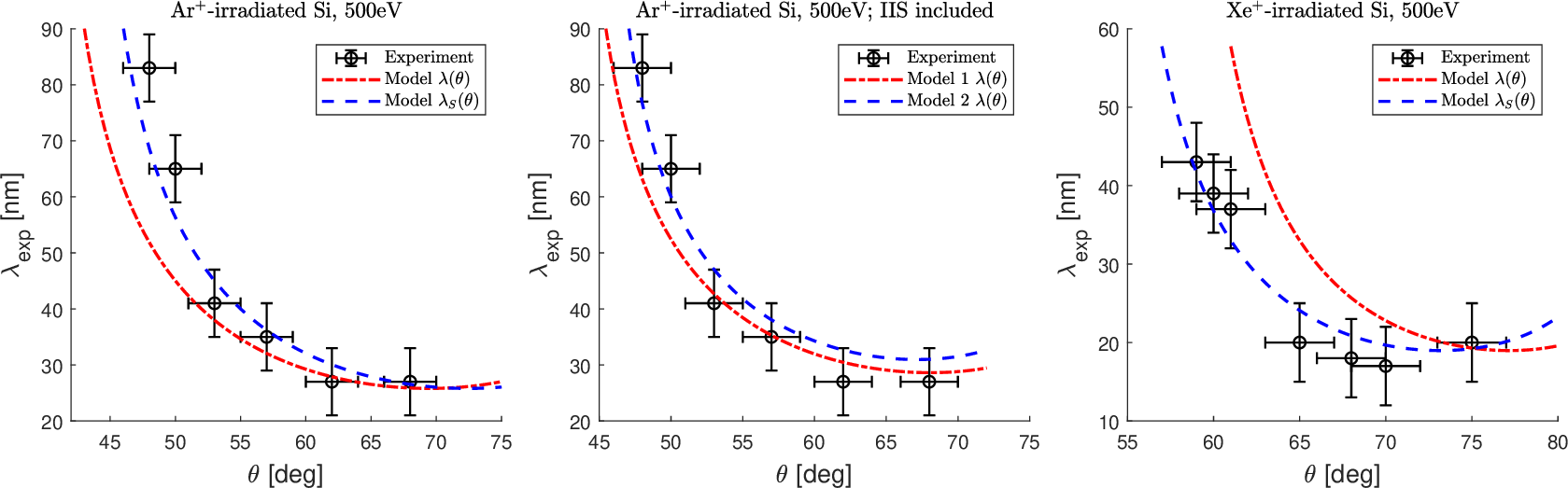}
\end{center}
\caption{Comparison of Equation (\ref{wavelengths}), Equation (\ref{wavelengths_withIIS}, and experimental data for 500eV Ar$^+$-irradiated Si and 500eV Xe$^+$-irradiated Si. Experimental data are due to \cite{moreno-barrado-etal-PRB-2015}. In all panels, we use $fA_D = 3\times 10^{-4}$ 1/s and $\eta = 150$ GPa s, as estimated in \cite{norris-etal-SREP-2017} \textbf{Left:} The red dash-dotted line, $\lambda(\theta)$, is a theoretical prediction from Equation (\ref{wavelengths}). The blue dashed line, $\lambda_S(\theta)$, is the same prediction as the red line, but with $\theta$ shifted by 3 degrees for the sake of comparison. \textbf{Middle:} The red dash-dotted line, $\lambda(\theta)$, is a theoretical prediction from Equation (\ref{wavelengths_withIIS}) $fA_I=6\times10^{-3}$ 1/s used as a fit parameter to increase the theoretical $\theta_c$ to match the experimental $\theta_c \approx 46^{\circ}$. The blue dashed line uses $fA_I = 1\times10^{-2}$ 1/s instead in order to illustrate the $\theta_c$-shifting effect of IIS. \textbf{Right:} A comparison similar to Figure \ref{fig:MBfigure}(a), but for Xe$^+$-irradiated Si at 500eV. The red dash-dotted curve is the model prediction based on Equation (\ref{wavelengths}), $fA_D = 3\times10^{-4}$ 1/s and $\eta = 150$ GPa s with no IIS. The blue dashed curve is a shift in $\theta$ of the red dash-dotted curve to correct for $\theta_c$. Up to this shift, the shape of $\theta_c$ is apparently correct, including the noticeable increase in wavelength past 70$^{\circ}$. }
\label{fig:MBfigure}
\end{figure}

In Figure \ref{fig:MBfigure}, we compare the experimental wavelength measurements of \cite{moreno-barrado-etal-PRB-2015} for 500eV Ar$^+$- and Xe$^+$- irradiated Si using Equation (\ref{wavelengths}). We use parameters $fA_D = 3 \times 10^{-4}$ 1/s and $\eta = 150$ GPa s, as estimated in \cite{norris-etal-SREP-2017} for 1000eV Ar$^+$-irradiated Si at a lower flux, $a=3.6, \alpha=0.6, \beta=0.9$ for Xe$^+$ irradiated Si at 500eV, and $a=2.5, \alpha=1, \beta=1.1$ for Ar$^+$ irradiated Si at 500eV. For Ar$^+$-irradiated Si at 500eV, Equation (\ref{wavelengths}) predicts $\theta_c \approx 39^{\circ}$, in contrast with the experimentally-observed $\theta_c \approx 46^{\circ}$. However, it has been shown elsewhere that ion-induced swelling (IIS) and the effect of phase-change at the amorphous-crystalline interface can each exert a stabilizing effect, tending to increase $\theta_c$ while simply shifting $\lambda(\theta)$. Instead of (\ref{wavelengths}), we might consider

\begin{equation}
\lambda(\theta) = \sqrt{\frac{2\gamma h_0(\theta)}{-3fA_D\eta\big(\cos(2\theta)\big(3\tau_{00} +2h_0\tau_{01}\big) - \sin(2\theta)\big(\frac{3}{2}\tau_{10} + h_0\tau_{11} + 3\tau_{00}\frac{x_0}{h_0} + 3\tau_{01}x_0\big) \big) - \frac{3fA_I\cos(\theta)}{2}}}, \label{wavelengths_withIIS}
\end{equation}
a plausible functional form if IIS were included in the present model, based on \cite{evans-norris-JPCM-2023}. A similar term, albeit with different angular dependence, results from the consideration of phase change at the amorphous-crystalline interface \cite{evans-norris-JEM-2024}. The effect of including $-\frac{fA_I\cos(\theta)}{2}$ is considered in the center panel of Figure \ref{fig:MBfigure}. Adding a small amount of IIS to the model, we shift the predicted wavelengths to higher $\theta$, achieving good alignment with experimental observations without affecting the shape of the theoretical $\lambda(\theta)$. That is: Equation (\ref{wavelengths}) predicts the correct shape of $\lambda(\theta)$, and the correct magnitudes, but they occur too early with respect to $\theta$. In contrast, Equation (\ref{wavelengths}) predicts $\theta_c$ larger than experimentally-observed $\theta_c$ for 500eV Xe$^+$-irradiated Si. Again permitting $\lambda(\theta)$ to shift with respect to $\theta$, we find that Equation (\ref{wavelengths}) can be aligned with experimental data for 500eV Xe$^+$-irradiated Si. In particular, the present model appears to correctly capture the shape of $\lambda(\theta)$ as $\theta$ approaches grazing incidence, where a noticeable increase in $\lambda$ is observed.

\subsection{600eV Ar\texorpdfstring{$^+$}{+}-irradiated Si: wavelength evolution and flux dependence}
\begin{figure}[h!]
\begin{center}
\includegraphics[width=1\linewidth]{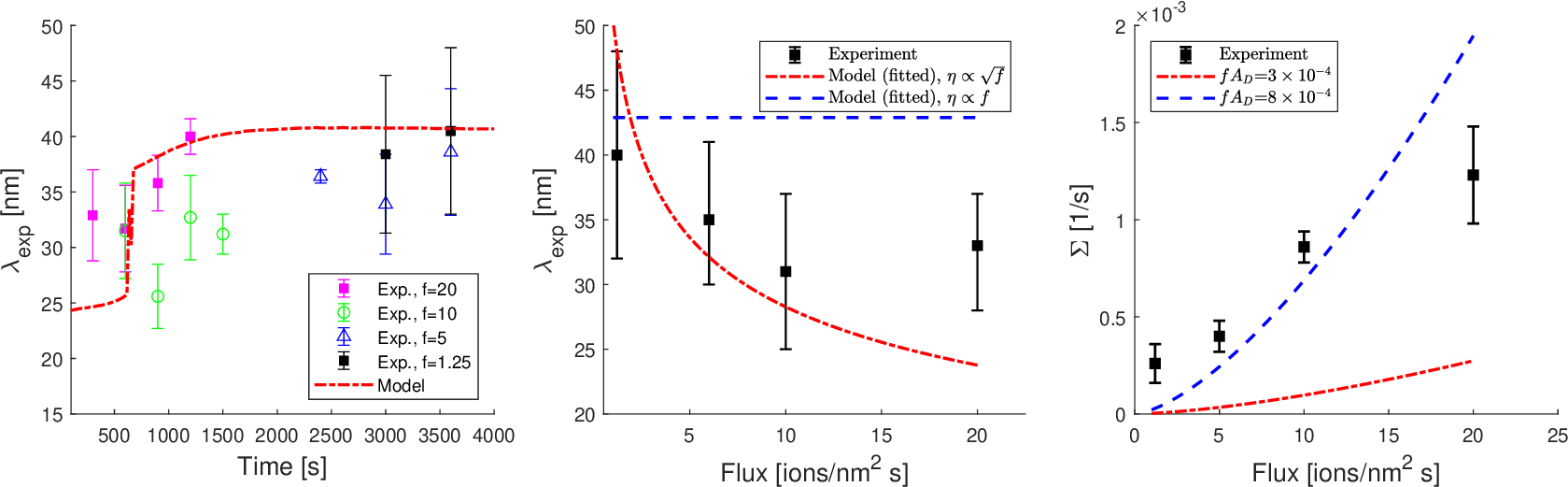}
\end{center}
\caption{\textbf{Left:} Comparison of the wavelength evolution of simulated 600eV Ar$^+$-irradiated Si based on Equation (\ref{NLPDE}) and $58^{\circ}$ irradiation with the experimental data of \cite{ishii-thesis-2013} at $68^{\circ}$. Fluxes $f$ are shown in ions/nm$^2$/s. We scale the simulated time to compare with the experimental time series. \textbf{Middle:} A qualitative comparison of the flux dependence of $\lambda(68^{\circ})$ based on Equation (\ref{wavelengths}) and experimental observations. $A_D$ and $\eta^{-1}(f=1)$ are used as fit parameters. The red dash-dotted line assumes $\eta^{-1} \propto \sqrt{f}$ and the blue dashed line assumes $\eta^{-1} \propto f$. \textbf{Right:} Comparison of experimental and theoretical growth rate $\Sigma$ (amplification factor in \cite{ishii-thesis-2013}) based on Equation (\ref{amplification_factor}).}
\label{fig:Ishiifigure}
\end{figure}

In Figure \ref{fig:Ishiifigure}, we consider the morphological evolution of the irradiated semiconductor surface based on Equation (\ref{NLPDE}) for 600eV Ar$^+$-irradiated Si. As before, we use $fA_D = 3\times 10^{-4}$1/s and $\eta = 150$ GPa s \cite{norris-etal-SREP-2017}. Based on BCA simulations \cite{ziegler-biersack-littmark-1985-SRIM}, we estimate $a=2.7, \alpha=1.2, \beta=1.2$ for this experimental system. To facilitate comparison between Equation (\ref{NLPDE}) and the time series of \cite{ishii-thesis-2013}, we have kept the times reported by \cite{ishii-thesis-2013} and scaled time in the time series produced by Equation (\ref{NLPDE}) by a factor of 600. Since the experiments of \cite{norris-etal-SREP-2017} were conducted at 0.02 ions/nm$^2$/s, this might suggest comparison with experimental flux around 12 ions/nm$^2$/s. Indeed, of the fluxes used in \cite{ishii-thesis-2013}, our simulations of (\ref{NLPDE}) appear to coincide best with the experimental flux of 20 ions/nm$^2$/s.

In \cite{ishii-thesis-2013}, flux-dependence of $\lambda(68^{\circ})$ is observed. Comparing with our theoretical wavelength estimate (\ref{wavelengths}), we find that obtaining the correct flux-scaling of $\lambda$ depends on assuming $\eta^{-1} \propto \sqrt{f}$, as in \cite{ishii-etal-JMR-2014,evans-heyen-PRE-2026}, rather than $\eta^{-1} \propto f$, as is commonly assumed \cite{norris-etal-NCOMM-2011,hofsass-APA-2015,norris-etal-SREP-2017}. For higher values of flux, we attribute the increase in $\lambda$ to thermal effects: as seen in \cite{evans-heyen-PRE-2026}, higher fluxes can plausibly reduce fluidity $\eta^{-1}$ by speeding up recombination of Frenkel pairs, leading to higher $\eta$, hence higher $\lambda$. In Figure \ref{fig:Ishiifigure}(a), we compare the time-evolution of experimental $\lambda(68^{\circ})$ with simulated $\lambda(58^{\circ})$, observing that the present model predicts $\theta_c\approx 38$, in contrast with experimental $\theta_c\approx 48$. We find reasonable agreement between model and experimental observations. In particular, simulated wavelength evolution appears to produce the correct saturated wavelength, $\lambda \approx 40$ nm.

We also find that Equations (\ref{amplification_factor}) and (\ref{wavelengths}) reproduce the form of the flux dependence observed for wavelength $\lambda(\theta)$ and growth rate $\Sigma$ for 68$^{\circ}$ irradiation, respectively. However, $fA_D = 3\times 10^{-4}$ 1/s is evidently too small to produce agreement with experimentally observed $\Sigma$. We plot $fA_D =8\times10^{-4}$ 1/s for comparison as an apparent best-fit value.


\subsection{1000eV Kr\texorpdfstring{$^+$}{+}-irradiated Si: ripple velocity and wavelengths}
\begin{figure}[h!]
\begin{center}
\includegraphics[width=.66\linewidth]{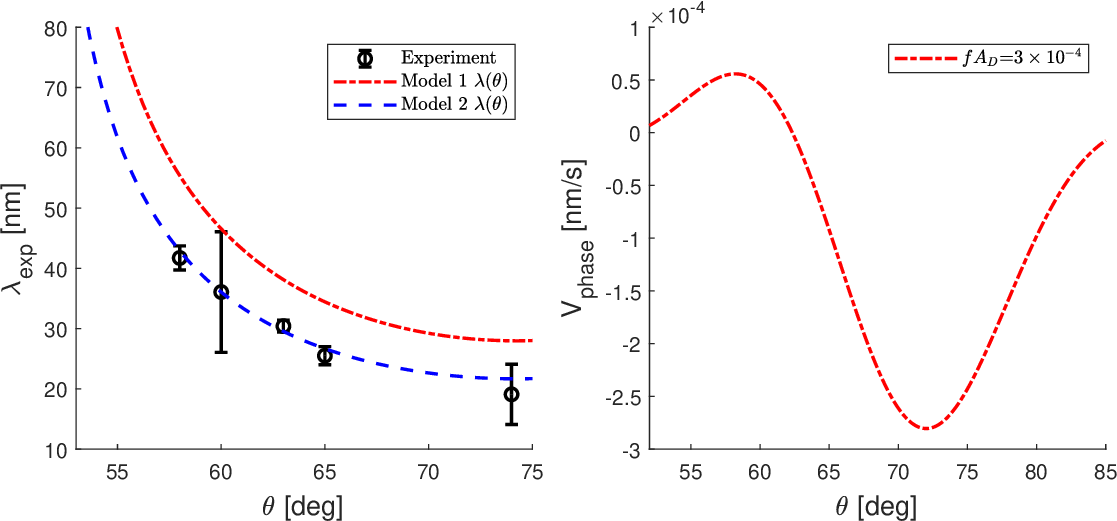}
\end{center}
\caption{\textbf{Left:} Comparison of Equation (\ref{wavelengths}) to the 1keV Kr$^+$-irradiated Si data due to \cite{myint-ludwig-PRB-2021-Kr-bombardment} and the 2keV Kr$^+$-irradiated Si data of \cite{seo-et-al-JPCM-2022} when rescaled by $\sqrt{2}$, consistent with the empirical scaling $\lambda(E) \propto \sqrt{E}$ for a fixed ion-target-angle combination, as noted in, e.g., \cite{hofsass-bobes-zhang-JAP-2016}. The red dash-dotted line uses $fA_D=3\times 10^{-4}$ 1/s and $\eta=150$  GPa s, while the blue dashed line uses $fA_D= 3\times 10^{-4}$ 1/s and $\eta=250$ GPa s. \textbf{Right}: Phase velocity of the fastest-growing ripple based on Equation (\ref{velocity}). While the predicted velocities are too small by several orders of magnitude when compared with the experiments of \cite{myint-ludwig-PRB-2021-Kr-bombardment}, this curve correctly predicts ripple translation into the projected ion-beam direction (``upstream") at the experimentally studied irradiation angle, $\theta=65^{\circ}$. }
\label{fig:1keVKrSifigure}
\end{figure}

There is seemingly only one system for which we can compare the present model directly with experimental observations of ripple velocity. For 65 degrees irradiation of Si by Kr at 1keV, \cite{myint-ludwig-PRB-2021-Kr-bombardment} reports observing ripple velocities about an order of magnitude greater than expected from an erosion-based theory \cite{bradley-harper-JVST-1988,bradley-PRB-2011b}, and into the projected ion beam direction. They measure a ripple velocity of 4.8 nm/s and an observed wavelength $\lambda(65^{\circ}) \approx 24-27$nm depending on irradiation time. This is enough to estimate $fA_D$ and $\eta$ for the given experimental system based on Equation (\ref{velocity}).

It is also desirable to compare with experimental wavelength data. However, there seems to be only very little available for this particular experimental system. On the other hand, a trend that $\lambda \propto \sqrt{E}$ for a fixed ion-target combination and irradiation angle has been widely noted; see, e.g., \cite{hofsass-bobes-zhang-JAP-2016}. Hence, to obtain approximate experimental wavelength data for comparison with Equation (\ref{wavelengths}), we take the $\lambda(\theta)$ measurements of \cite{engler-etal-PRB-2014,seo-et-al-JPCM-2022} for 2keV Kr$^+$-irradiated Si, divide each by $\sqrt{2}$, and combine them with the $\lambda(65^{\circ}) \approx 24-27$ nm reported by \cite{myint-ludwig-PRB-2021-Kr-bombardment} for 1keV Kr$^+$-irradiated Si. This combined set of experimental wavelength data is shown in Figure \ref{fig:1keVKrSifigure}; we also note the apparent consistency of these two data sets. While Equation (\ref{wavelengths}) equipped with $fA_D = 3\times 10^{-4}$ 1/s and $\eta = 150$ GPa s (the red curve) is an imperfect fit, a slightly higher $\eta = 250$ GPa s brings the model into excellent alignment with the experimental data. Given the large uncertainties involved in estimating $\eta$ \cite{volkert-JAP-1991,ishii-etal-JMR-2014,perkinsonthesis2017,norris-etal-SREP-2017}, it is interesting that an adjustment well-within an order of magnitude so greatly improves agreement with experimental data.

Curiously, the parameter values used to obtain good agreement with $\lambda(\theta)$ appear to be irreconcilable with the reported ripple velocity by several orders of magnitude. Within the present model, this would suggest the presence of a missing mechanism which contributes exclusively, or almost exclusively, to the imaginary part of the amplification rate, while only very weakly to the real part. This point is revisited in the Discussion.

\color{black}
\subsection{2000eV Kr\texorpdfstring{$^+$}{+}-irradiated Si: roughening and wavelength evolution}
\begin{figure}[h!]
\begin{center}
\includegraphics[width=1\linewidth]{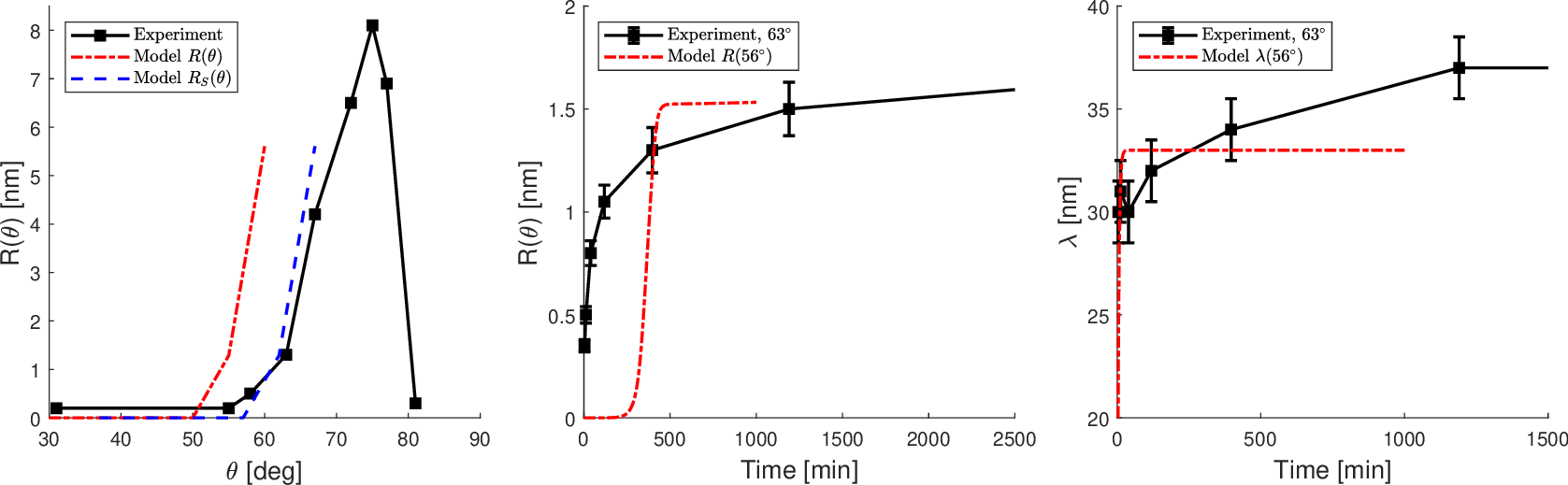}
\end{center}
\caption{Comparison of the experimental data of \cite{engler-etal-PRB-2014} with predictions of Equation (\ref{NLPDE}) using $fA_D = 10^{-3}$ 1/s and $\eta=500$ GPa s and $a,\alpha,\beta$ consistent with 2keV Kr$^+$-irradiated Si. \textbf{Left:} Angle-dependent saturated roughness. The red dash-dotted curve is the simulation result and the blue dashed curve is the same result shifted by 7 degrees. \textbf{Middle:} Comparison of experimental roughness data (black boxes with error bars) and numerical integration of Equation (\ref{NLPDE}) (red dash-dotted curve). \textbf{Right:} Comparison of experimental wavelength data (black boxes with error bars) and numerical integration of Equation (\ref{NLPDE}) (red dash-dotted curve). Notably, in contrast with Figure \ref{fig:Ishiifigure}, we have \textit{not} rescaled time to facilitate comparison, suggesting that our parameter values may be close to realistic. }
\label{fig:2keVKrSifigure}
\end{figure}

Here, we consider the experiments of \cite{engler-etal-PRB-2014} for 2keV Kr$^+$-irradiated Si. Several time series are reported, including of roughness $R$ and wavelength $\lambda$ at 63 and 72 degrees irradiation. We find that the parameters $fA_D = 3\times 10^{-4}$ 1/s and $\eta=150$ GPa poorly account for the data; in particular, they predict an extremely small amplification factor, leading to prohibitively long run times for the model to saturate. However, we can use Equation (\ref{NLPDE}) and the parameters $fA_D = 10^{-3}$ 1/s and $\eta = 500$ GPa s to reproduce the angle dependent saturated roughness and, broadly, the time series shown for 63 degrees irradiation. As before, we find that the present model predicts $\theta_c$ lower than experimental --- here, by about seven degrees (see Figure \ref{fig:2keVKrSifigure} (left panel)). Hence we consider the comparison of Equation (\ref{NLPDE}) where we simulate irradiation at 56 degrees in order to compare with experiments conducted at 63 degrees. We find broadly favorable comparison, particularly with the angle-dependent saturation of roughness, where the shape of the theoretical $R(\theta)$ --- up to a translation in $\theta$ --- agrees almost exactly with experimental observations. On the other hand, the time series for surface roughness and wavelength predicted by Equation (\ref{NLPDE}) is rather different: experimental observations show slower growth and later saturation.


While \cite{engler-etal-PRB-2014} also presents time series data for Kr$^+$-irradiated Si at 2keV and $\theta=72^{\circ}$, we attempt no comparison with Equation (\ref{NLPDE}). Since the present model assumes small slopes, one does not expect that surfaces of large roughness, such as those that develop for near-grazing irradiation angles, would be well-modeled.

In comparing the present model with experimental data for 2keV Kr$^+$-irradiated Si, we have departed somewhat from established parameter values. Nonetheless, it is intriguing that the apparent ``best-fit" parameter values, $fA_D = 10^{-3}$ 1/s and $\eta=500$ GPa s, are each well within an order of magnitude of those observed in \cite{norris-etal-SREP-2017}, and very comfortably within the range of parameter estimates obtained elsewhere, or used to explain experimental observations in low-energy semiconductor irradiation \cite{van-dillen-etal-APL-2001-colloidal-ellipsoids,van-dillen-etal-APL-2003-colloidal-ellipsoids,van-dillen-etal-PRB-2005-viscoelastic-model,vandillen-etal-prb-2006,george-etal-JAP-2010,norris-etal-NCOMM-2011,moreno-barrado-etal-PRB-2015,perkinsonthesis2017,munoz-garcia-etal-PRB-2019,evans-norris-JPCM-2023,evans-norris-JEM-2024}.

\section{Discussion}
\paragraph{Summary.} In the present work, we have studied a continuum model of irradiation-induced nano-patterning of a semiconductor surface. Following previous work, we have treated the first few nanometers of the irradiated semiconductor --- the amorphous layer --- as a fluid of extremely high viscosity. The amorphous layer is separated from the underlying crystalline wafer (e.g., silicon or germanium) by the amorphous-crystalline interface. We have obtained closed-form expressions for the depth-dependence of ion implantation, or ``local flux", by an asymptotic analysis. Within the same calculation, we obtain the shape of the amorphous-crystalline interface in terms of the level sets of implanted ion density through a nearly-flat free surface perturbed by a sinusoid of small amplitude and arbitrary wavenumber. This constitutes an improvement on the work of \cite{evans-norris-JEM-2024}, which was restricted to long-wave perturbations of the macroscopically flat free interface.

These results in hand, our continuum model leads to a generalized Kuramoto-Sivashinsky equation (gKS) describing the evolution of the irradiated surface under the physical hypothesis that a parcel of the amorphous layer undergoes anisotropic plastic flow of an intensity proportional to the instantaneous, local flux experienced by that parcel. The coefficients of the gKS are explicitly connected to this physical hypothesis, three easily-obtained ion implantation parameters ($a,\alpha,\beta$), and two other quantities --- $A_D$ and $\eta$ --- which are, at least in principle, experimentally accessible, and have already been estimated for some systems; see, for example, \cite{madi-thesis-2011,norris-etal-NCOMM-2011,ishii-etal-JMR-2014,norris-etal-SREP-2017,perkinsonthesis2017,evans-norris-JPCM-2023}.

From our gKS, we obtain numerous theoretical predictions: (i) the angle-dependent wavelength $\lambda(\theta)$, (ii) minimal angle $\theta_c$ at which pattern formation initiates, (iii) amplification factor $\Sigma$, (iv) ripple velocity, (v) flux-dependence of $\lambda$, (vi) angle-dependent saturated surface roughness $R(\theta)$, and (vii) predicted time-series of surface evolution. Of these, (i)-(iv) require only a linear stability study of our gKS, while the latter (v)-(vii) require the numerical integration of the full nonlinear PDE.

\paragraph{Contributions.} We have compared our theoretical model to experimental observations for several experimental systems of interest, and several forms of available experimental data: angle-dependent wavelengths $\lambda(\theta)$ for 500eV Ar$^+$-irradiated Si and 500eV Xe$^+$-irradiated Si; flux-dependent amplification rate and wavelengths, and time series wavelength data for 600eV Ar$^+$-irradiated Si; angle-dependent wavelengths and ripple velocity for 1keV Kr$^+$-irradiated Si; and angle-dependent roughness and time series of roughness and wavelength for 2keV Kr$^+$-irradiated Si. For lack of experimental parameter estimates, we simply use the existing estimates $fA_D = 3\times 10^{-4}$ 1/s and $\eta = 150$ GPa s, based on \cite{norris-etal-SREP-2017} for 1000eV Ar$^+$-irradiated Si at a flux of 0.02 ions/nm$^2$/s.

Interestingly, these parameters lead to reasonable agreement across a wide variety of systems. We find that when $\theta_c$ is predicted by our model to be erroneously higher or lower than experimental results, most of the dynamics are the same --- or very similar --- up to the value of the shift in $\theta$. This shift in $\theta$ without affecting the underlying dynamics can be plausibly explained by the presence of IIS or BA, which we have excluded from the present model for simplicity, and which we have shown have the effect on $\lambda(\theta)$ of simply inducing an offset. However, including IIS would have led to two coupled nonlinear partial differential equations of gKS-type: too much for a first study.

The primary contribution of the present work is in the development of a model evidently capable of reconciling some experimental observations with theory equipped only with two free parameters --- $A_D$ and $\eta$ --- restricted to values within an order of magnitude of known estimates. However, several other contributions of various degrees of importance are summarized as follows:
\begin{itemize}
    \item The gKS that we obtain contains its own $(h_{xx})^2$ term. This term, sometimes referred to as a Kuramoto-Sivashinsky-type nonlinearity, did not emerge from the analysis of \cite{seo-et-al-PRB-2025}, which produces a gKS otherwise similar in form. The fact that such a term emerges here from a completely viscous-flow model is an intriguing difference. The simulations of \cite{seo-et-al-PRB-2025} seem to imply that the present model, when extended from the $x$-axis to the $x,y$ plane, might produce the raised triangular features shown in, e.g., \cite{carter-etal-REDS-1977,seo-et-al-PRB-2025} for non-negligible sputtering scenarios.
    \item The present model tends to produce the correct $\lambda(\theta)$, possibly up to a shift in $\theta$, and correctly identifies that Ar$^+$-irradiated Si should have a lower $\theta_c$ than Xe$^+$-irradiated Si. That this occurs even for the same parameter values $fA_D$ and $\eta$ suggests that the physical hypothesis of coupling the strength of APF to the local ion flux broadly captures the correct physics. This may motivate the development of a more accurate approximation of Equation (\ref{Dsmallcurvature}) than the simple linearizations used here.
    \item Our asymptotic approximations, both of the amorphous-crystalline interface under various assumptions about the free interface and of the depth-dependence of ion implantation, or local flux, are already novel. They can be used in their given form even if the present model is ultimately refuted by further experimental work.
\end{itemize}

\paragraph{Future work.} Perhaps the most striking disagreement between the present model and experimental observations is on ripple velocity. While our model correctly predicts ripple propagation into the direction of the ion beam (i.e., ``upstream") for 1keV Kr$^+$-irradiated Si, as well as reasonable agreement with experimental wavelengths, particularly when we hypothesize a somewhat higher $\eta$, the velocity predicted by the present model is several orders of magnitude too slow compared with what is reported by \cite{myint-ludwig-PRB-2021-Kr-bombardment}. This suggests a missing component of the present model which would contribute strongly to the imaginary part of the amplification rate while only very weakly to the real part. This missing component is not ion-induced swelling or interfacial phase change, since those two mechanisms, while excluded here, are known to contribute negligibly to the imaginary part for long-wave perturbations of the free surface \cite{Swenson_2018,evans-norris-JPCM-2022,evans-norris-JPCM-2023,evans-norris-JEM-2024}. Measurements of ripple velocity, similar to \cite{myint-ludwig-PRB-2021-Kr-bombardment}, for other systems would also be illustrative.

The present model also produces a slope distribution on the roughness-saturated surface different than those observed by \cite{engler-etal-PRB-2014}, although it can reasonably well-reproduce those observed by \cite{seo-et-al-PRB-2025} at low fluences. To the extent that one believes that an APF or ion-hammering-like mechanism is responsible for evolving surface roughness, wavelength and $\theta_c$ selection, it still appears that the slopes are selected by some other process, possibly erosion acting alongside viscous-flow mechanisms. This provides an interesting area for future study.

Nonetheless, including IIS and BA would be natural next steps for the present work --- in particular, because we have now seen that a model based on APF can produce qualitative and near-quantitative accuracy for some aspects of morphological evolution when supplemented by the physical hypothesis that its intensity varies locally like the pointwise flux.

It will also be important to reconcile the present work with that of \cite{li-et-al-NIMB-2026}, where MOSS experiments for 250eV Ar$^+$-irradiated Si appear to show that pattern formation and stress evolution are independent. It appears likely that reconciling the apparent explanatory power of the present model with the observations of \cite{li-et-al-NIMB-2026} will depend on developing a fundamental understanding of what causes APF --- or a phenomenologically similar behavior --- in low-energy irradiated semiconductors. It will also be important to reconcile the present modeling approach with the erosive-redistributive framework.

\section*{Acknowledgments}
TPE gratefully acknowledges the generous support of the National Science Foundation through DMS-1840260 while at Southern Methodist University and DMS-2136198 while at University of Utah. 

\appendix
\section{Asymptotic approximations}
For readability, we have suppressed most mathematical details in the main text. Here, we obtain expressions for the instantaneous ion dose given a prescribed free surface. First, we consider a free surface perturbed by a sinusoid of small amplitude and a single wavenumber. From this, we obtain the deposition profile in the irradiated bulk, and we consider level sets of the deposition field in order to obtain a theoretical description of the amorphous-crystalline boundary. These results constitute a generalization of our own earlier work, \cite{evans-norris-JEM-2024}. A similar approach is then applied in the case of small curvature and small slopes for perturbations of non-vanishing amplitude, and the results are used in obtaining the nonlinear partial differential equation (\ref{NLPDE}).

\subsection{Free surface perturbed by small-amplitude sinusoid}
\paragraph{Small-amplitude expansion with geometric flux dilution.}
While (\ref{deposition-integral}) cannot be evaluated in closed-form for an arbitrary surface $h(X,t)$, it \textit{can} be approximated in the limit of sinusoidal perturbations of a single, arbitrary wavenumber $k$ and amplitude $\epsilon$ to an otherwise flat surface --- that is,
\begin{equation}
    h(X,t) = \epsilon \tilde{h}_1e^{\Sigma t + ikX}.
\end{equation}
Expanding the integrand of (\ref{deposition-integral}) in $\epsilon$ and integrating at $\mathcal{O}(1)$ and $\mathcal{O}(\epsilon)$, we obtain the approximation
\begin{equation}
	D_f(x,z) = D_{f0}(z) + \epsilon \tilde{h}_1\exp\big(\Sigma t + ikx \big)D_{f1}(z) + \mathcal{O}(\epsilon^2), \label{deposition-linearized-withflux}
\end{equation}
where
\begin{equation}
    D_{f0}(z) = \frac{\cos(\theta)}{\sqrt{2\pi(\alpha^2c^2 + \beta^2s^2)}}\exp\bigg(-\frac{(z+ac)^2}{2(\alpha^2c^2 + \beta^2s^2)} \bigg)
\end{equation}
and
\begin{equation}
\begin{gathered}
D_{f1}(z) = \frac{\cos(\theta)}{2\sqrt{\pi A}\alpha^3\beta^3}\exp\bigg(\frac{\tilde{B}^2(z)}{4A}-C(z) \bigg)\bigg[c_1(z)- c_2\frac{\tilde{B}(z)}{2A} \bigg].
\end{gathered}
\end{equation}
In the above,
\begin{equation*}
	\begin{gathered}
		c_1(z) = z\left( \alpha^2\sin^2(\theta) + \beta^2\cos^2(\theta)\right) + \cos(\theta)a\beta^2 + ik\alpha^2\beta^2\tan(\theta) \\
		c_2 = \cos(\theta)\sin(\theta)(\alpha^2-\beta^2),
	\end{gathered}
\end{equation*}
and
\begin{equation*}
\begin{gathered}
A = \frac{\beta^2s^2 + \alpha^2c^2}{2\alpha^2\beta^2 }; \hspace{.25cm}
B(z) = \frac{zsc(\alpha^2-\beta^2) - a\beta^2s}{\alpha^2\beta^2}; \\
\tilde{B}(z) = \frac{zsc(\alpha^2-\beta^2) - a\beta^2s}{\alpha^2\beta^2} + ik; \hspace{.25cm}
C(z) = z^2\frac{\beta^2c^2+\alpha^2s^2}{2\alpha^2\beta^2} + z\frac{ac}{\alpha^2} + \frac{a^2}{2\alpha^2}.
\end{gathered}
\end{equation*}
where $c=\cos(\theta)$ and $s=\sin(\theta)$.

\paragraph{Small-amplitude expansion without geometric flux dilution.} Revisiting the calculation above where there is no flux dilution, we obtain instead
\begin{equation}
	D(x,z) = D_0(z) + \epsilon \tilde{h}_1\exp\big(\Sigma t + ikx \big)D_1(z) + \mathcal{O}(\epsilon^2), \label{deposition-linearized-noflux}
\end{equation}
where all dependence on $x$ is carried by the leading exponential at $\mathcal{O}(\epsilon)$, and all dependence on $z$ is described by
\begin{equation}
    D_0(z) = \frac{1}{\sqrt{2\pi(\alpha^2c^2 + \beta^2s^2)}}\exp\bigg(-\frac{(z+ac)^2}{2(\alpha^2c^2 + \beta^2s^2)} \bigg),
\end{equation}
which is simply a Gaussian in the depth $z$, and
\begin{equation}
\begin{gathered}
    D_1(z) = \frac{1}{\alpha^2\beta^2\sqrt{2\pi(\alpha^2c^2 + \beta^2s^2)}}\exp\bigg(\frac{\tilde{B}(z)^2 }{4A} - C(z)\bigg)\times \\
    \bigg[z(\alpha^2s^2 + \beta^2s^2) + ca\beta^2 - cs(\alpha^2-\beta^2)\frac{\big(zsc(\alpha^2-\beta^2)-a\beta^2s + ik\alpha^2\beta^2\big)}{(\alpha^2c^2 + \beta^2s^2) }\bigg].
\end{gathered}
\end{equation}



\paragraph{Amorphous-crystalline interface in small-amplitudes limit.} Proceeding from the linearized deposition field (\ref{deposition-linearized-noflux}), we express the depth as
\begin{equation}
	z= g_0 + \epsilon \tilde{g}_1\exp\big(\Sigma t + ikx\big),
\end{equation}
where we solve for $g_0$ and $\tilde{g}_1$ such that
\begin{equation}
	D(x,z) = D_0(z) + \epsilon \tilde{h}_1\exp\big(\Sigma t + ikx \big)D_1(z) + \mathcal{O}(\epsilon^2) = D_c,
\end{equation}
where $D_c$ is some critical threshold beyond which amorphization proceeds. Hence, we asymptotically approximate the level curve where the instantaneous ion dose is $D_c$. We first establish $D_c$ by computing the first and second central moments, $M_1$ and $M_2$, of the linearized deposition field. By definition,
\begin{equation}
\begin{gathered}
	M_1 = \int_{-\infty}^{\infty}z\big(D_0(z) + \epsilon \tilde{h}_1\exp\big(\Sigma t + ikx\big)D_1(z)\big)dz, \\
	M_2 = \int_{-\infty}^{\infty}z^2\big(D_0(z) + \epsilon \tilde{h}_1\exp\big(\Sigma t + ikx\big)D_1(z)\big)dz,
\end{gathered}
\end{equation}
and we adopt the notation
\begin{equation}
\begin{gathered}
    M_1 = M_{10} + \epsilon \tilde{h}_1\exp\big(\Sigma t + ikx \big)M_{11} \\
    M_2 = M_{20} + \epsilon \tilde{h}_1\exp\big(\Sigma t + ikx \big)M_{21}.
\end{gathered}
\end{equation}
Then we define $D_c = D(x,z=\mu -2\sigma)$, so that the amorphization threshold is assigned as two standard deviations $\sigma$ away from the mean $\mu$ along the $z$-axis and \textit{into} the irradiated substrate, where $\mu = M_1$ and $\sigma = \sqrt{M_2 - M_1^2}$. Expanding in $\epsilon$, we compute
\begin{equation}
\begin{gathered}
    \mu = \mu_{0} + \epsilon \tilde{h}_1\exp\big(\Sigma t + ikx \big)\mu_1, \\
    \mu_0 = M_{10}, \hspace{.5cm} \mu_1 = M_{11},
\end{gathered}
\end{equation}
and
\begin{equation}
\begin{gathered}
    \sigma = \sigma_{0} + \epsilon \tilde{h}_1\exp\big(\Sigma t + ikx \big)\sigma_1, \\
    \sigma_0 = \sqrt{M_{20} - M_{10}^2}, \hspace{.5cm} \sigma_1 = \frac{M_{21}-2M_{10}M_{11}}{2\sqrt{M_{20} - M_{10}^2}}.
\end{gathered}
\end{equation}
Then we set
\begin{equation}
\begin{gathered}
    D_0(g_0 + \epsilon g_1) + \epsilon h_1 D_1(g_0+\epsilon g_1) = \\ D_0(\mu_0 + \epsilon\mu_1 - 2(\sigma_0 + \epsilon \sigma_1)) + \epsilon h_1D_1(\mu_0 + \epsilon\mu_1 - 2(\sigma_0 + \epsilon \sigma_1)).
\end{gathered}
\end{equation}
Upon linearization in $\epsilon$, we obtain at $\mathcal{O}(1)$
\begin{equation}
    D_0(g_0) = D_0(\mu_0 - 2\sigma_0), \label{leadingorderg}
\end{equation}
and restricting the domain of interest to the plane \textit{below} $z=\mu_0$ immediately implies\footnote{We note that the symmetry of the Gaussian ellipsoid describing implantation into the bulk permits, in principle, two solutions to Equation \ref{leadingorderg}, and we are only interested in solving for the lower one. Restriction to the lower half-plane restores invertibility.} $g_0 = \mu_0 - 2\sigma_0$. At $\mathcal{O}(\epsilon)$,
\begin{equation}
\begin{gathered}
    \frac{\partial D_0}{\partial z}|_{z=g_0}\tilde{g}_1 + D_1(g_0)\tilde{h}_1 = \\ \tilde{h}_1\frac{\partial D_0}{\partial z}|_{z=\mu_0 - 2\sigma_0}\cdot\big(M_{11} - \frac{M_{21}-2M_{10}M_{11}}{\sqrt{M_{20} - M_{10}^2}}\big) + \tilde{h}_1D_1\big(M_{10} - 2\sqrt{M_{20}-M_{10}^2}\big).
\end{gathered}
\end{equation}
Then
\begin{equation}
    \frac{\tilde{g}_1}{\tilde{h}_1} = \frac{\frac{\partial D_0}{\partial z}|_{z=\mu_0 - 2\sigma_0}\cdot\big(M_{11} - \frac{M_{21}-2M_{10}M_{11}}{\sqrt{M_{20} - M_{10}^2}}\big) + D_1\big(M_{10} - 2\sqrt{M_{20}-M_{10}^2}\big) - D_1(g_0)}{\frac{\partial D_0}{\partial z}|_{z=g_0}}.
\end{equation}
This simplifies to
\begin{equation}
    \frac{\tilde{g}_1}{\tilde{h}_1} = M_{11} - \frac{M_{21}-2M_{10}M_{11}}{\sqrt{M_{20} - M_{10}^2}} \equiv \mu_1 - 2\sigma_1. \label{g1h1-momentform}
\end{equation}
Hence $\frac{\tilde{g}_1}{\tilde{h}_1}$ is expressible \textit{entirely} in terms of the moments of the deposition field. However, previous work \cite{cuerno-etal-NIMB-2011,castro-cuerno-ASS-2012,castro-etal-PRB-2012,norris-PRB-2012-linear-viscous,moreno-barrado-etal-PRB-2015,Swenson_2018,munoz-garcia-etal-PRB-2019,evans-norris-JPCM-2022,evans-norris-JPCM-2023,evans-norris-JEM-2024} typically uses a mean film thickness $h_0(\theta)$ measured against an amorphous-crystalline interface at $z\approx 0$ and lateral phase shift $x_0(\theta)$ between the sinusoidal boundaries. In particular, the lateral phase shift is commonly incorporated into models as 
\begin{equation}
    \frac{\tilde{g}_1}{\tilde{h}_1} = \exp(-ikx_0),
\end{equation}
as in \cite{moreno-barrado-etal-PRB-2015,Swenson_2018,evans-norris-JPCM-2022,evans-norris-JPCM-2023,evans-norris-JEM-2024}. We must now extract $x_0$ from Equation (\ref{g1h1-momentform}). In the setting of complex arithmetic,
\begin{equation}
    a+bi = re^{i\psi}
\end{equation}
where $a,b,r,\psi\in \mathbb{R}$, $i$ is the imaginary unit, and $\psi$ is the polar angle in the complex plane. This leads us to equate $x_0 = -\frac{\psi}{k}$, and the factor $r$ that appears will determine the \textit{relative amplitude} between the free and amorphous-crystalline interfaces. Then
\begin{equation}
    \text{Re}\bigg(M_{11} - \frac{M_{21}-2M_{10}M_{11}}{\sqrt{M_{20} - M_{10}^2}} \bigg) + i\text{Im}\bigg(M_{11} - \frac{M_{21}-2M_{10}M_{11}}{\sqrt{M_{20} - M_{10}^2}} \bigg) = r_0e^{i\psi}
\end{equation}
where
\begin{equation}
    r_0 = \sqrt{ \bigg[\text{Re}\bigg(M_{11} - \frac{M_{21}-2M_{10}M_{11}}{\sqrt{M_{20} - M_{10}^2}} \bigg)\bigg]^2 + \bigg[\text{Im}\bigg(M_{11} - \frac{M_{21}-2M_{10}M_{11}}{\sqrt{M_{20} - M_{10}^2}} \bigg)\bigg]^2},
\end{equation}
\begin{equation}
    \psi = \arctan\Bigg[\frac{\text{Im}\big(M_{11} - \frac{M_{21}-2M_{10}M_{11}}{\sqrt{M_{20} - M_{10}^2}} \big)}{\text{Re}\big(M_{11} - \frac{M_{21}-2M_{10}M_{11}}{\sqrt{M_{20} - M_{10}^2}} \big)}\Bigg],
\end{equation}
and simplifying leads to
\begin{equation}
    x_0 = -\frac{1}{k}\arctan\Bigg[\frac{\text{Im}\big(M_{11} - \frac{M_{21}-2M_{10}M_{11}}{\sqrt{M_{20} - M_{10}^2}} \big)}{\text{Re}\big(M_{11} - \frac{M_{21}-2M_{10}M_{11}}{\sqrt{M_{20} - M_{10}^2}} \big)}\Bigg]. \label{x0_moments}
\end{equation}
Now the linearized amorphous-crystalline interface $g$ is fully characterized by the moments of the deposition field $M_{10}, M_{11}, M_{20}, M_{21}$, and we write $$\frac{\tilde{g}_1}{\tilde{h}_1} = r_0\exp(-ikx_0).$$
The above quantity appears repeatedly throughout linear stability analyses of hydrodynamic-type models of ion-induced pattern formation \cite{norris-PRB-2012-linear-viscous,moreno-barrado-etal-PRB-2015,swenson-thesis-2018,evans-norris-JPCM-2022,evans-norris-JPCM-2023,evans-norris-JEM-2024}.

\paragraph{Amorphous-crystalline interface in physical parameters.} Above, calculations were carried out formally in terms of the moments $M_{10}, M_{11}, M_{20}, M_{21}$. Here, we record $h_0, x_0$ and $r_0$ in terms of the basic ion implantation statistics $a,\alpha,\beta$. After simplification, we have
\begin{equation}
\begin{gathered}
    M_{10} = -ac, \\
    M_{20} = c^2a^2 + (\alpha^2c^2 + \beta^2s^2), \\
    M_{11} = \exp\bigg(-\frac{k^2}{4}\big( (\alpha^2+\beta^2)-(\alpha^2-\beta^2)\cos(2\theta)\big)\bigg)e^{-iaks}, \\
    M_{21} = -2M_{11}c\big(a - ik(\alpha^2-\beta^2)s\big),
\end{gathered}
\end{equation}
leading to
\begin{equation}
    M_{11} - \frac{M_{21}-2M_{10}M_{11}}{\sqrt{M_{20} - M_{10}^2}} = M_{11}\bigg[1 - \frac{2ik(\alpha^2-\beta^2)sc}{\sqrt{\alpha^2c^2 + \beta^2s^2} } \bigg].
\end{equation}
As in the main text, $c=\cos(\theta)$ and $s=\sin(\theta)$. Simplifying the expressions for $M_{10}, M_{20}, M_{11},$ and $M_{21}$ finally yields
\begin{equation}
\begin{gathered}
    h_0(\theta) = -g_0(\theta) = ac + 2\sqrt{\alpha^2c^2 + \beta^2s^2}, \\
    x_0(\theta;k) = \frac{1}{k}\arctan\bigg[\frac{\sin(ak\sin(\theta)) + \frac{2k(\alpha^2-\beta^2)sc\cos(ak\sin(\theta))}{\sqrt{\alpha^2c^2+\beta^2s^2} } } {\cos(ak\sin(\theta)) - \frac{2k(\alpha^2-\beta^2)sc\sin(ak\sin(\theta))}{\sqrt{\alpha^2c^2+\beta^2s^2 } } } \bigg],   \\
    r_0(\theta;k) = e^{-\frac{k^2}{4}\big((\alpha^2+\beta^2) - (\alpha^2-\beta^2)\cos(2\theta) \big)} \bigg[1 + \frac{k^2(\alpha^2-\beta^2)^2\sin^2(2\theta)}{\alpha^2c^2 + \beta^2s^2}\bigg].
\end{gathered}
\end{equation}
In the above, we have written $h_0(\theta) = -g_0(\theta)$, representing a change of reference: in the preceding calculations, the upper interface was taken as $z\approx 0$, and we derived the location of the lower interface, $z=g_0(\theta)$. Reverting to the more widely used convention, we subtract $g_0(\theta)$ from the height $z$ of each of the interfaces and then identify $z=-g_0(\theta)$ with the more-conventional $h_0(\theta)$. In Figure \ref{fig:CCbased}, $h_0(\theta),x_0(\theta;k)$ and $r_0(\theta;k)$ are plotted for 1000eV Ar$^+$-irradiated Si.

\paragraph{Note on discontinuities in $x_0(\theta;k)$.} The function $x_0(\theta;k)$ is intended to return the \textit{positive} phase shift between a sinusoidal free interface of wavenumber $k$, say, $h(x)=\sin(kx)$ and the corresponding sinsuoidal amorphous-crystalline interface, say, $g(x)=\sin\big(k(x-x_0)\big)$. However, when 
\begin{equation}
\cos(ak\sin(\theta)) - \frac{2k(\alpha^2-\beta^2)sc\sin(ak\sin(\theta))}{\sqrt{\alpha^2c^2+\beta^2s^2 } }=0 \label{singularities}
\end{equation}
for some $\theta$, $x_0(\theta;k)$ experiences a jump continuity, leading to non-physical output. To restore the intended physical meaning, one instead computes
\begin{equation}
    x_0(\theta;k) = \frac{1}{k}\bigg(\arctan\bigg[\frac{\sin(ak\sin(\theta)) + \frac{2k(\alpha^2-\beta^2)sc\cos(ak\sin(\theta))}{\sqrt{\alpha^2c^2+\beta^2s^2} } } {\cos(ak\sin(\theta)) - \frac{2k(\alpha^2-\beta^2)sc\sin(ak\sin(\theta))}{\sqrt{\alpha^2c^2+\beta^2s^2 } } } \bigg] + \pi\sum_{j=1}^{n}\text{He}(\theta-\theta^*_j)\bigg)
\end{equation}
where $\text{He}(x)$ is the Heaviside function and $\theta^*_j$ is the j$^{\text{th}}$ root $\theta$ of (\ref{singularities}) between $0^{\circ}$ and $90^{\circ}$ for fixed parameters $a,\alpha,\beta,k$. It appears that for \textit{most} physically-realistic parameter combinations ($a,\alpha,\beta,k$) (\ref{singularities}) has no such roots. In particular, as $k\to0$, (\ref{singularities}) has no solutions $\theta$ for any $(a,\alpha,\beta)$. We address the matter here only for mathematical completeness.

\paragraph{Special limits: $k\to 0, k\to \infty$.} It is clear that $h_0(\theta)$ is independent of wavenumber $k$, and is therefore expected to be suitable for any perturbation. It is also identical to the result obtained using the argument of \cite{evans-norris-JEM-2024}. For arbitrary wavenumber, $x_0(\theta;k)$, on the other hand, depends on $k$. Using L'H\^{o}pital's (Bernoulli's) Rule, we recover the $k \to 0$ limit,
\begin{equation}
    x_0(\theta;0) = a\sin(\theta) + \frac{2(\alpha^2-\beta^2)\sin(\theta)\cos(\theta) }{\sqrt{\alpha^2\cos^2(\theta) + \beta^2\sin^2(\theta)}} \label{interfaces-x0-longwave}
\end{equation}
as expected. Again for $k \to 0$, $r_0(\theta;k) \to 1$, as expected (since this term does not appear in the analysis of \cite{evans-norris-JEM-2024}). Finally, when $k \to \infty$, both $r_0(\theta;k) \to 0$ and $x_0(\theta;k) \to 0$. This means that the amorphous-crystalline boundary is flat regardless of the shape of the free interface, and the notion of a phase-shift between the free and crystalline-amorphous interfaces loses meaning. We note, however, that the $k \to \infty$ case is essentially nonphysical, as the period of the free interface cannot physically be smaller than that of the lateral distance occupied by two bonded substrate atoms, and extremely large values of $k$ already invalidate the continuum assumption underlying all hydrodynamic-type models. This observation is included, again, for mathematical completeness.

\subsection{Free surface of small curvature}
Now we are interested in developing an approximation of Equation (\ref{deposition-integral}) valid for free surfaces of small curvature but, possibly, non-vanishing amplitude and not of a single wavenumber. The small-curvature approximation can then be used to obtain a further simplification in the case of small slopes. Both are expected to be of value for obtaining weakly nonlinear evolution equations for the surface, as in \cite{oron-davis-bankoff-RMP-1997,craster-matar-APS-2009,ajaev-book-2012,munoz-garcia-etal-PRB-2019}.

We first observe that not all ions contribute equally to the value of $D(x,z,t)$ at a fixed but arbitrary point $(x,z)$, denoting the $X$-axis domain of influence $[x_L,x_R]$. This domain can be related to the chosen point $(x,z)$ in the bulk as follows: $x-x_L$ must be the distance from $x$ so that the most distant ellipsoidal implantation region deposits less than 5\% of its ions at $(x,z)$. Observe that $\sigma = 2\sqrt{\alpha^2s^2 + \beta^2c^2}$ gives the $X$-axis extent of the 95$^{\text{th}}$-percentile ellipsoid in line with the collision cascade's center, whose $X$-axis coordinate is $x_L + as$. As special cases, we note that $\sigma$ correctly reproduces the limits $\sigma \to 2\alpha$ when $\theta \to \frac{\pi}{2}$ and $\sigma \to 2\beta$ when $\theta \to 0$. Then the leftmost $X$ coordinate through which an ion could enter through $z=h(x,t)$ and be deposited with a probability of at least 5\%  at $(x,z)$ is $x_L = x - as - 2\sqrt{\alpha^2s^2 + \beta^2c^2}$. A similar argument produces $x_R = x - as + 2\sqrt{\alpha^2s^2 + \beta^2c^2}.$ Loosely, this says that the radius of curvature of the irradiated surface be small compared to the collision cascade's standard deviation. 

If the slope of $h(X,t)$ changes slowly enough over the interval $(x_L,x_R)$ that the surface admits a linear approximation
\begin{equation}
    h(X,t) = h(x,t) + (X-x)h_x
\end{equation}
over that interval, then a small-curvature approximation of $D_f(x,z,t)$ is obtained by carrying out the integral
\begin{equation}
\begin{gathered}
    D_f(x,z,t) \approx \frac{\cos(\theta) + h_x\sin(\theta)}{\sqrt{1+h_x^2}}\int_{-\infty}^{+\infty} \Psi\big(x,z;X,h(x,t) + (X-x)h_x\big)dX \\
    = \frac{1}{2\pi\alpha \beta} \frac{c + h_xs}{\sqrt{1+h_x^2}} \int_{-\infty}^{+\infty} e^{\big(-\frac{[(x-X)s - (z-[h(x,t) + (X-x)h_x] )c-a]^2}{2\alpha^2} - \frac{[(x-X)c + (z-[h(x,t) + (X-x)h_x])s]^2}{2\beta^2} \big)}dX.    
\end{gathered}
\end{equation}
Now the Gaussian integral above can be computed exactly using standard techniques to obtain the closed-form approximation
\begin{equation}
    D_f(x,z,t) \approx \frac{1}{\sqrt{2\pi}} \frac{c + h_xs}{\sqrt{1+h_x^2}} \frac{\exp\bigg[-\frac{\big((z-h(x,t) ) + a(c+sh_x)\big)^2 }{2\beta^2(s-ch_x)^2 + 2\alpha^2(c+sh_x)^2} \bigg]}{\sqrt{\beta^2(s-ch_x)^2 + \alpha^2(c+sh_x)^2}}.
\end{equation}
Here, we note that all explicit $x$-dependence has been absorbed into $h(x,t)$. Moreover, if all slopes $h_x$ of interest are small in the laboratory coordinates, then (\ref{Dsmallcurvature}) admits a series expansion in $h_x \approx 0$. In the small slope expansion, we recover the result of \cite{evans-norris-JEM-2024}, again finding
\begin{equation}
\begin{gathered}
    g(x,t) = h(x-x_0(\theta),t) - \big(ac + 2\sqrt{\alpha^2c^2+\beta^2s^2}\big), \\ 
    x_0(\theta) = a\sin(\theta) + \frac{2(\alpha^2-\beta^2)\sin(\theta)\cos(\theta) }{\sqrt{\alpha^2\cos^2(\theta) + \beta^2\sin^2(\theta)}},
\end{gathered}
\end{equation}
as in \cite{evans-norris-JEM-2024}. We also note that linear or quadratic approximations of depth-dependence for ion-induced deformation feature in some of the literature \cite{moreno-barrado-etal-PRB-2015,munoz-garcia-etal-PRB-2019,evans-norris-JPCM-2023}. Following this, we record here the linear approximation of Equation (\ref{Dsmallcurvature}) about $z= h(x,t)$, $$D_f(x,z,t) \approx D_{f0}(x,z,t) + D_{f1}(x,z,t)\big(z-h(x,t)\big)$$ where
\begin{equation}
\begin{gathered}
    D_{f0}(x,z,t) = \frac{1}{\sqrt{2\pi}}\frac{c+h_xs}{\sqrt{1+h_x^2}}\frac{\exp\bigg[-\frac{\big(a(c+sh_x)\big)^2 }{2\beta^2(s-ch_x)^2 + 2\alpha^2(c+sh_x)^2} \bigg]}{\sqrt{\beta^2(s-ch_x)^2 + \alpha^2(c+sh_x)^2}} \\
    D_{f1}(x,z,t) = \frac{-a(c+h_xs)D_{f0}(x,z,t)}{2sch_x(\alpha^2-\beta^2) + s^2(h_x^2\alpha^2 + \beta^2) + c^2(h_x^2\beta^2 + \alpha^2 )}.
\end{gathered}
\end{equation}


\section{Derivation of the nonlinear PDE}
 Continuing from the main text, we apply the standard lubrication scalings \cite{oron-davis-bankoff-RMP-1997,craster-matar-APS-2009,ajaev-book-2012}. These scalings and nondimensionalization are applied simultaneously as $x = \frac{h_0 \tilde{x}}{\epsilon}, y=\frac{h_0 \tilde{y}}{\epsilon}, z = h_0\tilde{z}, t = \frac{h_0 \tilde{t}}{\epsilon u_0}, u = u_0\tilde{u}, v=u_0\tilde{v}, w=\epsilon u_0 \tilde{w}$. Here,
$h_0$ is again the angle-dependent film thickness, $u_0$ is mean lateral flow velocity, and $\epsilon$ is a typical, nondimensionalized wave number.  
We also nondimensionalize the pressure, surface energy, upper and lower interfaces and ion flux as $p = \frac{\eta u_0 \tilde{p}}{\epsilon h_0}, \gamma = \frac{\eta u_0 \tilde{\gamma}}{\epsilon^3}, h = h_0\tilde{h}, g = h_0\tilde{g}$ and $fA_D = \frac{u_0 \tilde{f}\tilde{A_D}}{h_0}$. At leading order, and after restricting attention to the $x$ axis, we have the following. In the interior,
\begin{gather}
	\tilde{p}_{0,\tilde{x}} = \tilde{u}_{0,\tilde{z},\tilde{z}} + 2\tilde{f}\tilde{A_D}\big(D_{11}\tau_{D,\tilde{x}} + D_{13}\tau_{D,\tilde{z}}\big) \\
	\tilde{p}_{0,\tilde{z}} = 2\tilde{f}\tilde{A_D}D_{33}\tau_{D,\tilde{z}} \\
	\tilde{u}_{0,\tilde{x}} + \tilde{w}_{0,\tilde{z}} = 0.
\end{gather}
At $\tilde{z} = \tilde{g}$,
\begin{gather}
	\tilde{u}_{0} = \tilde{w}_{0} =0
\end{gather}
and at $\tilde{z} = \tilde{h}$,
\begin{gather}
	-\tilde{h}_{\tilde{x}}\big(-\tilde{p}_{0} - 2\tilde{f}\tilde{A_D}\tau_D\tilde{D}_{11}\big) + \tilde{u}_{0,\tilde{z}} - 2\tilde{f}\tilde{A_D}\tau_D\tilde{D}_{13} = \tilde{\gamma}\tilde{h}_{\tilde{x}}\tilde{h}_{\tilde{x}\tilde{x}}  \\
	-\tilde{p}_{0} - 2\tilde{f}\tilde{A_D}\tau_{D}\tilde{D}_{33} = \tilde{\gamma}\tilde{h}_{\tilde{x}\tilde{x}}.
\end{gather}
It is straightforward to solve for $\tilde{p}_0, \tilde{u}_0$ and $\tilde{w}_0$ by noticing that $\tilde{p}_0$ can be obtained up to an unknown function of $x$ by integrating $\tilde{p}_{0,\tilde{z}} = 2\tilde{f}\tilde{A_D}D_{33}\tau_{\tilde{z}}$. Then the second boundary condition at $\tilde{z}=\tilde{h}$ is applied to fully determine $\tilde{p}_0$. Next, $\tilde{u}_0$ is found by carrying out two integrations in $\tilde{z}$ of the first momentum balance equation. The two resulting unknown functions of $\tilde{x}$ are determined by using the boundary conditions for $\tilde{u}_0$ at $\tilde{z}=\tilde{g}$ and $\tilde{z}=\tilde{h}$. We obtain $\tilde{w}_0$ by integrating conservation of mass and applying the boundary condition at $\tilde{z}=\tilde{g}$. Finally, we substitute into the kinematic condition
\begin{gather}
	\tilde{h}_{\tilde{t}} = -\tilde{u}_0\tilde{h}_{\tilde{x}} + \tilde{w}_0,
\end{gather}
and restore dimensional quantities while assigning $g(x,t) =h(x-x_0(\theta),t) - h_0(\theta) \approx h(x,t)-x_0(\theta)h_x(x,t)-h_0(\theta)$, as in (\ref{interfaces-x0-longwave}). Next, retaining terms up to $\mathcal{O}(\epsilon^3)$, we find 
\begin{equation}
\begin{gathered}
    h_t = C_1h_{xx} + C_2h_xh_{xx} + C_3h_x^2h_{xx} + C_4h_{xx}^2 + C_{5}h_{xxx} + C_{6}h_{x}h_{xxx} \\ + C_{7}h_{xx}h_{xxx} + C_{8}h_{xxxx},
\end{gathered}
\end{equation}
which describes the temporal evolution of the height field in terms of its own derivatives --- in particular, a nonlinear partial differential of generalized Kuramoto-Sivashinsky type. For readability, we have suppressed the arguments of, e.g., $C_{1} = C_1(\theta,a,\alpha,\beta,f,A,\gamma,\eta)$. We record these coefficients $C_{1}-C_{8}$ in the Appendix. The PDE above contains many of of the same nonlinearities considered in \cite{pearson-bradley-JPCM-2015,gelfand-bradley-PLA-2015,perkinson-etal-JPCM-2018,munoz-garcia-etal-PRB-2019}, where interesting parameter-dependent dynamical behavior was observed, including (i) symmetry-breaking \cite{pearson-bradley-JPCM-2015,gelfand-bradley-PLA-2015}, (ii) coarsening \cite{pearson-bradley-JPCM-2015} and saturation of roughness \cite{munoz-garcia-etal-PRB-2019}, and (iii) angle-dependent reversal of the velocity \cite{munoz-garcia-etal-PRB-2019}. At the same time, some nonlinearities in the above \textit{do not} appear in \cite{munoz-garcia-etal-PRB-2019}, an alternative model of ion-induced nano-patterning, suggesting nontrivial differences in dynamics.

\section{Coefficients of the nonlinear PDE}
The coefficients of the nonlinear PDE are:
\begin{equation*}
\begin{gathered}
    C_1 = fA_Dh_0^2\bigg[ \cos(2\theta)\big(3\tau_{00} +2h_0\tau_{01}\big) - \sin(2\theta)\big(\frac{3}{2}\tau_{10} + h_0\tau_{11} + 3\tau_{00}\frac{x_0}{h_0} + 3\tau_{01}x_0\big)\bigg], \\
    C_2 = fA_Dh_0^2\bigg[\cos(2\theta)\bigg(6\tau_{10} + 4h_0\tau_{11} + 12\frac{x_0}{h_0}\tau_{00} + 12\tau_{01}x_0\bigg) \\ - \sin(2\theta)\bigg(6\tau_{10}\frac{x_0}{h_0} + 6\tau_{11}x_0 + 3\tau_{00}\frac{x_0^2}{h_0^2} + 6\tau_{01}\frac{x_0^2}{h_0}\bigg)\bigg],  \\
    C_3 = fA_Dh_0^2\bigg[\cos(2\theta)\big(18\tau_{10}\frac{x_0}{h_0} + 18\tau_{11}x_0 + 9\tau_{00}\frac{x_0^2}{h_0^2} + 18\tau_{01}\frac{x_0^2}{h_0}\big) \\ - \sin(2\theta)\big(\frac{9}{2}\tau_{10}\frac{x_0^2}{h_0^2} + 9\tau_{11}\frac{x_0^2}{h_0} + 3\tau_{01}\frac{x_0^3}{h_0^2}\big)\bigg] \\
    C_4 = fA_D\bigg[2h_0^2\tau_{10}x_0 + 3h_0^3\tau_{11}x_0\cos(2\theta)\bigg] - \frac{h_0^2\gamma}{\eta} \\
\end{gathered}
\end{equation*}
\begin{equation*}
\begin{gathered}
    C_{5} = fA_D\bigg[\frac{2}{3}h_0^3\tau_{10} + \frac{3}{4}h_0^4\tau_{11}\cos(2\theta)\bigg] \\ 
    C_{6} = fA_D\bigg[ 3h_0^3x_0\tau_{11}\cos(2\theta) + 2h_0^2x_0\tau_{10}\bigg] - \frac{h_0^2\gamma}{\eta} \\
\end{gathered}
\end{equation*}
\begin{equation*}
\begin{gathered}
    C_{7} = \frac{-h_0^2x_0\gamma}{\eta}; 
    C_{8} = \frac{-h_0^3\gamma}{3\eta} \\
\end{gathered}
\end{equation*}
where $x_0,h_0,\tau_{00},\tau_{01},\tau_{10}$ and $\tau_{11}$ are functions of the geometric parameters $a,\alpha,\beta$ and global irradiation angle $\theta$, as described in the main text.

\bibliographystyle{plain}
\bibliography{bapaperrefs2}

\end{document}